\documentclass[journal ]{new-aiaa}
\usepackage[utf8]{inputenc}
\usepackage{textcomp}

\usepackage{graphicx}
\usepackage{amsmath}
\usepackage[version=4]{mhchem}
\usepackage[separate-uncertainty=true]{siunitx}
\usepackage{longtable,tabularx,booktabs}
\usepackage{subfig}
\usepackage{color}

\title{Spectral Features for Re-entry Break-up Event Identification}

\author{David Leiser \footnote{PhD Student, HEFDiG, Institute of Space Systems, Pfaffenwaldring 29, 70569 Stuttgart, Germany, AIAA Member.} and Stefan Loehle\footnote{Group Leader, HEFDiG, Institute of Space Systems, Pfaffenwaldring 29, 70569 Stuttgart, AIAA Senior Member}}
\affil{High Enthalpy Flow Diagnostics Group, Institute of Space Systems, University of Stuttgart, Germany}
\author{Stefanos Fasoulas \footnote{Director, Institute of Space Systems, Pfaffenwaldring 29, 70569 Stuttgart, Germany, AIAA Senior Member.}}
\affil{Institute of Space Systems, University of Stuttgart, Germany}
\begin{document}

\maketitle

\begin{abstract}
The fragmentation of two aerospace aluminum alloys is investigated in a ground testing facility including mechanical loads as occurring due to aerodynamic forces in a real atmospheric entry event at three trajectory points. The emission spectroscopic analysis shows that these materials fail after distinct alkali metal features are observed in the spectra. The two alloys feature characteristic emissions of the different alkali metals. The presence of lithium lines that have previously been exclusively attributed to battery failure in observation campaigns may be considered as a marker for aluminum breakup. This is particularly interesting for future entry observations because it allows a new insight into the structural failure processes of the demising spacecraft. The lack of emission of alloying elements points to these spectra being a candidate for the determination of spacecraft demise. The identification of such features in ground testing will allow a more certain identification of specific break-up events.
\end{abstract}

% \section*{Nomenclature}

% \noindent(Nomenclature entries should have the units identified)

% {\renewcommand\arraystretch{1.0}
% \noindent\begin{longtable*}{@{}l @{\quad=\quad} l@{}}
% $A$  & amplitude of oscillation \\
% \end{longtable*}}

\section{Introduction}
		% General Background
\lettrine{T}{he} missing understanding of the disintegration of spacecraft structures during the atmospheric entry flight is the main driving parameter for the calculation of ground impact risk. Additionally, the full demise of the spacecraft becomes increasingly important, because of the rapidly increasing number of Low Earth Orbit (LEO) which undergo uncontrolled entry after the mission~\cite{ESA-SD_2021_01}. The recently launched satellite systems such as Starlink and OneWeb with thousands of satellites have furthermore only short lifetimes of 3-5\,years which again increases the amount of entering space debris. It is of utmost interest for the space industry to predict the re-entry and demise accurately and space debris problems are a main topic of the European Space Agency under the Space Debris Initiative~\cite{ESA-SD_2021_01}.
		% Specifc background
One option for the analysis of re-entry processes is the observation of spacecraft during re-entry which gives insight into the processes that dominate fragmentation and ultimately the demise of spacecraft~\cite{Loehle_2017_03}. Another option is to fly on-board systems which analyze the entry in-situ. However, this requires a comparably complex system and the hardware has to be sent to space. Four European Re-entry Break-up Recorders (REBR) were flown, of which three acquired data~\cite{Feistel_2013_01}. Finally, the experimental simulation of re-entry demise can be realized in ground testing facilities. In comparison with flight observations, this method allows investigating the particular features of an atmospheric entry leading to the full demise of spacecraft structures.
		% Statement of problem/Knowledge gap
The High Enthalpy Flow Diagnostics Group participated in almost all airborne re-entry observations using different spectroscopic instruments~\cite{Loehle_2010_03, Jenniskens_2010_03,	Loehle_2014_03,Jenniskens_2016_01}. We develop diagnostic methods to be applied in ground testing experiments~\cite{Leiser_2019_01} allowing us to assess the material processes in-situ. With a recently installed load cylinder, the simulation of mechanical forces during the aerothermal testing becomes available.
        % Here we show
In this study, material samples of the main structural components used in spacecraft were tested under combined aeromechanical and thermochemical loads.
        % Approach and Result
%During testing time resolved data is collected of the applied force and the resulting displacement as well as the plasma generator condition and the resulting surface temperature. Furthermore spectral information of the stagnation point was recorded and the surface reconstructed photogrammetrically. 
During testing the emission spectra of the stagnation point were observed by an Echelle spectrometer in \SIrange{250}{880}{\nano\meter}. The results of the present study show that depending on the mechanical stress and the aerothermal situation, the materials show different features in the spectral data.
% So what?
%Spectral features such as lithium lines that have previously been exclusively attributed to battery failure in observation campaigns may have to be considered as a marker for aluminum breakup. This is particularly interesting for future entry observations, because it allows a new insight into the structural failure processes of the demising spacecraft.

\section{Experimental Setup}
\begin{figure}[!ht]
	\centering
	\subfloat[Plasma wind tunnel PWK4.]{\includegraphics[trim= 100px 0px 0px 500px, clip,  height=7.0cm]{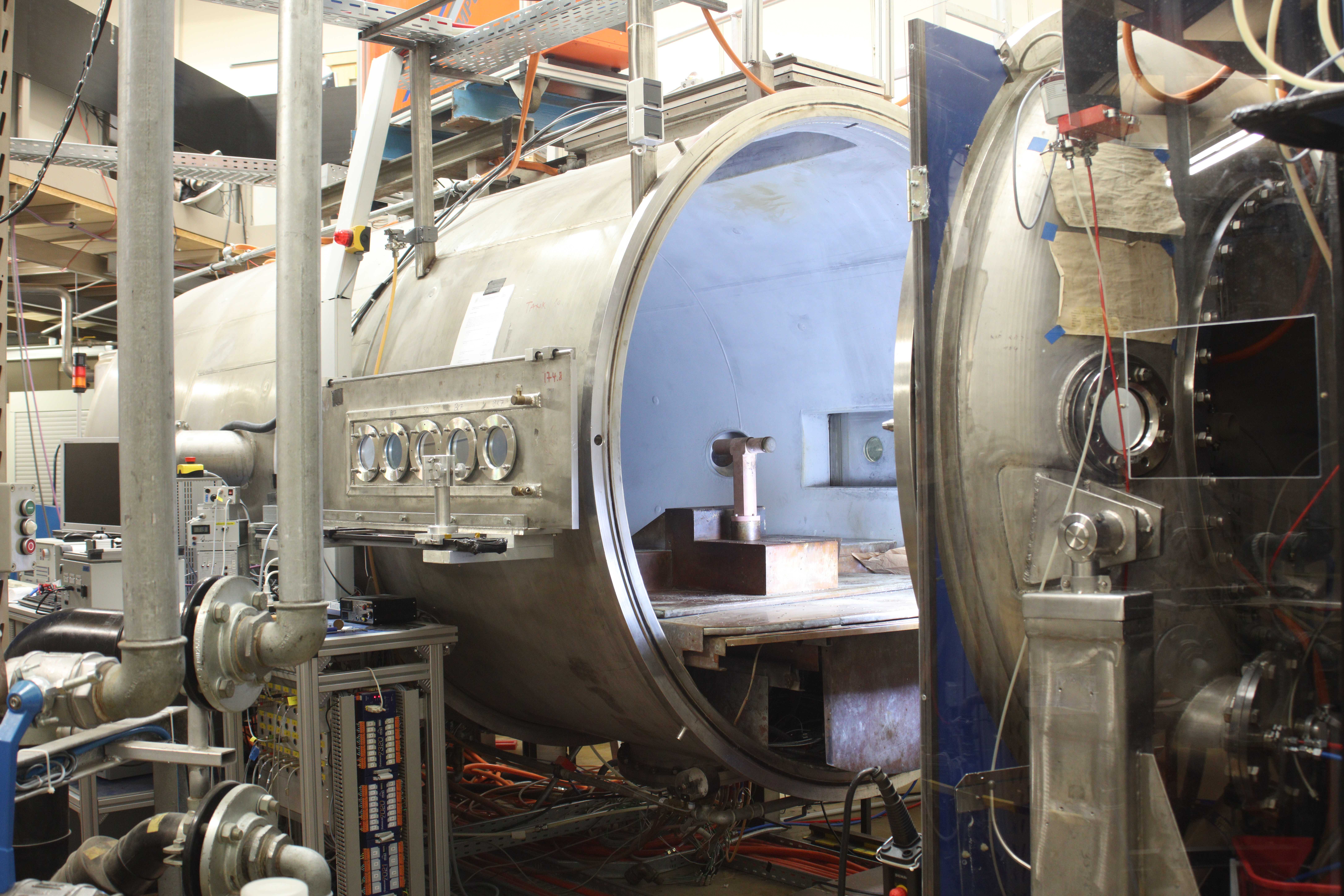}\label{fig:PWK_setup}}\quad
	\subfloat[Mechanical setup with sample.]{\includegraphics[trim= 500px 800px 500px 0px, clip, height=7.0cm]{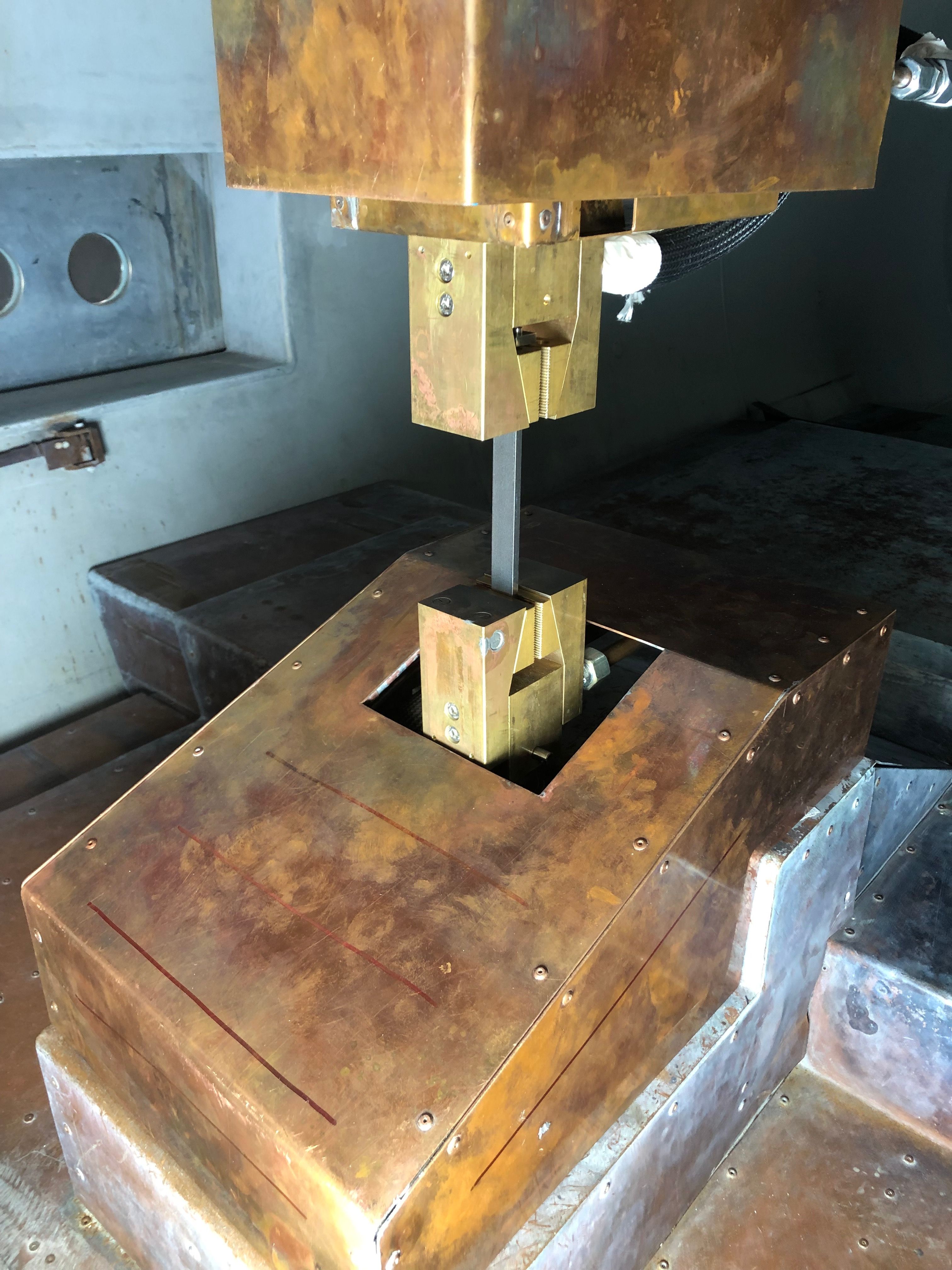}\label{fig:mech_setup}} 
	\caption{Test setup at IRS\label{fig:setup}}
\end{figure}

Experiments were conducted in the plasma wind tunnel facility PWK4 at the Institute of Space Systems - IRS~\cite{Loehle_2016_01} at the University of Stuttgart shown in Fig.~\ref{fig:PWK_setup}. The facility consists of a cylindrical vacuum vessel with a diameter of \SI{2}{\meter} and a length of \SI{6}{\meter}, connected to the central vacuum system with a four stage pump system that allows static pressures in the range of \SI{1}{\pascal}-\SI{50}{\kilo\pascal}. The plasma is generated by the thermal arc-jet plasma generator RB3, allowing for high local mass specific enthalpy at sufficiently high total pressures. The material samples are \SI{20}{\milli\meter} by \SI{5}{\milli\meter} flat bars with a length of \SI{90}{\milli\meter} mounted between a \SI{5}{\kilo\newton} electro-mechanical actuator by \emph{ZwickRoell} and the movable PWK test platform~\cite{Leiser_2019_01}. 

% \begin{figure}[ht!]
% 	\centering
% 	\includegraphics[width=.36\textwidth,height=.48\textwidth,trim= 10cm 30cm 10cm 0cm,clip]{images/190703_VerKritPWK_05.jpeg}
% 	\caption{Closeup view of the mechanical setup in PWK4.}
% 	\label{fig:mech_setup}
% \end{figure}

This setup is shown in Fig.~\ref{fig:mech_setup}, a flatbar sample is shown in the middle, held by the brass sample holders. The sensitive equipment housed in the plasma wind tunnel is protected from the aggressive environment by water-cooled copper shields with \SI{16}{\bar} high-pressure water, shown above and below the sample holders. %Special care was taken to minimize the contact cooling between the material samples and the interface points to water cooled elements such as the sample holders.
The samples are held by serrated grips manufactured out of steel. A pin allows the sample to be loaded with a pre-test clamping force which enables tests under minimal tensile force. The steel grips also functions as a thermal insulator between the samples and the watercooled brass sample holders. This minimizes the contact cooling of the samples.

During the facility startup the probe is positioned outside of the plasma flow. Once the nominal condition is set the probe is moved to the center of the flow.%, this marks the onset of the experiment and testime $t = \SI{0}{\second}$.

Two main types of tests were conducted, denoted as scenario 1 and 2. During scenario 1 tests, the samples are held in the plasma at minimal force $<\SI{10}{\newton}$ until the sample fails due to melting or it reaches an isothermal state . Scenario 2 tests are conducted by applying the nominally expected force as determined from the re-entry trajectory~\cite{Leiser_2021_01}. If the sample reaches an isothermal state in either scenario, the load is increased linearly until the sample fails. 

\subsection{PWK Flow Condition}
\begin{figure}[ht!]
	\centering
	\includegraphics[width=.6\textwidth]{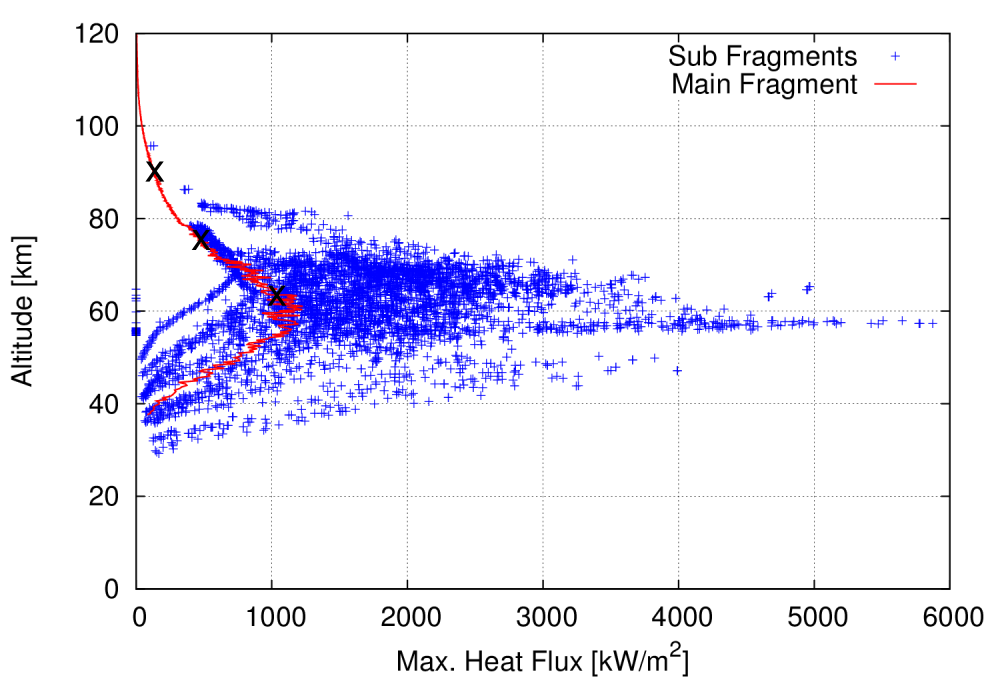}
	\caption{Re-entry fragmentation simulation of Sentinel-2 (courtesy HTG).}
	\label{fig:sentinel2}
\end{figure}

The flow conditions correspond to the re-entry trajectory of Sentinel-2~\cite{Lips_2020_01}, which falls into the ESA standard trajectory corridor~\cite{ESA-SD_2020_01, Beck_2020_01}, at an altitude of \num{65}, \num{75} and \SI{90}{\kilo\meter}. Figure~\ref{fig:sentinel2} shows the simulation and the chosen points for ground test simulation. These conditions were chosen as representative of three characteristic trajectory points on a re-entry from Low Earth Orbit~\cite{Koppenwallner_2005_01,Lips_2010_01}. 
The \SI{90}{\kilo\meter} point is early during re-entry and features small heat loads and the separation of outlying elements such as solar panels. \SI{75}{\kilo\meter} is around the break-up altitude with major fragmentation of the main structure of spacecraft. Peak heating occurs around \SI{65}{\kilo\meter}, this is a major factor in further structural break-up leading to subsystem release and determining the demise of components. 

The conditions necessary to obtain these trajectory points were determined by matching the heat flux and total pressure from flight to ground. 
%using local heat transfer simulation scaling~\cite{Leiser_2019_02}. For this the total enthalpy $h$, total presasure $p_{tot}$ and the Stanton number $St$ are matched from flight to ground testing. Transforming from axisymmetric re-entry bodies to planar test samples lets us express the Stanton number in terms of the more known boundary layer edge velocity gradient $\beta_e$. 
The tunnel conditions are characterized using a flat faced \SI{50}{\milli\meter} diameter axisymmetric total pressure and heat flux probe~\cite{Loehle_2016_06} machined from copper. The measured conditions are scaled to the aluminum flat bar samples using local heat transfer simulation scaling~\cite{Leiser_2019_02} and correcting for the difference in catalycity~\cite{diss_stoeckle}: 
 \begin{equation}
 \dot{q}_{\mathrm{flat bar}} = \dot{q}_{\mathrm{axisymmteric}}\sqrt{\frac{2.5}{4}}\label{eq:pl-axi}
 \end{equation}
The detailed approach for the translation of the flight situation to the flat bar testing in the ground testing facility are described in a different paper~\cite{Leiser_2020_01}. Thereby the condition on the flat bar equals the flight condition. The reference conditions, generator settings and resulting conditions are displayed in table~\ref{tab:generator_cond}.
\begin{table}[]
    \begin{center}
		\caption{Generator and flow condition corresponding with the trajectory points.}\label{tab:generator_cond}
		\renewcommand{\arraystretch}{1.2}
		\begin{tabular}{p{4cm}p{1.5cm}p{2cm}p{2cm}p{2cm}}
            \toprule
			Condition & & \SI{90}{\kilo\meter} & \SI{75}{\kilo\meter} & \SI{65}{\kilo\meter} \\
			\midrule	
			Generator conditions & Unit & & & \\
			\midrule	
			Arc current& \si{\ampere} & \num{350+-5} & \num{545+-4}&\num{500+-2}\\
			Arc voltage& \si{\volt} & \num{85+-2} & \num{90+-4}&\num{98+-2}\\

		    Mass flow rate & \si{\gram\per\second} & \num{4.00+-0.01} &\num{5.00+-0.01} &\num{6.52+-0.01} \\
			Ambient pressure& \si{\pascal} & \num{35+-1} & \num{201+-1} & \num{1250+-5}\\

			\midrule
			Test conditions &Unit&&&\\
			\midrule
			Reference distance& \si{\milli\meter} & \num{310+-0.1} & \num{310+-0.1} & \num{185+-0.1}\\
			Heat flux& \si{\kilo\watt\per\meter\squared} & \num{120+-2} & \num{497+-5} & \num{843+-5} \\
			Stagnation pressure& \si{\pascal} & \num{177+-2} & \num{1352+-4} & \num{4690+-10}\\
	        \bottomrule
		\end{tabular}
    \end{center}
\end{table}
%The heat flux given in table~\ref{tab:generator_cond} is the heat flux to the flat bar samples. The necessary conversion of the heat fluxes between axisymmetric and planar has been derived in previous publication. %The local mass specific enthalpy was calculated from the total pressure and heat flux using Marvin and Pope's formula~\cite{Marvin_1968_01}., the mach number was calculated using the Rayleigh-Pitot equation.

\subsection{Investigated Materials}

%Four materials were tested as the main structural constituents of most satellites, specifically ESA's ATV~\cite{Lips_2011_01}. It is expected that the main emission of radiation, aside from the air constituents, stems from the base or alloying elements of the samples.
Aluminum is the main element of most satellite structures, used for all manner of lightweight structures. Two well characterized alloys that are used extensively are Al-6060 and Al-7075~\cite{Lips_2011_01}. The alloying elements according to DIN EN 573-3~\cite{DIN573} are shown in table~\ref{tab:Alu_comp}
\begin{table}[ht]
	\begin{center}
		\caption{Aluminum alloys and their main alloying elements in mass percent.}\label{tab:Alu_comp}\vspace{1em}
		\renewcommand{\arraystretch}{1.2}
		\begin{tabular}{l|cccccccccc}
            & Al & Si & Fe & Cu & Mn & Mg & Cr & Zn & Ti & other\\
            \midrule
            Al-6060  & remainder & \numrange{0.3}{0.6} & \numrange{0.1}{0.3} & \num{0.1} & \num{0.1} & \numrange{0.35}{0.6} & \num{0.05} & \num{0.15} & \num{0.10} & \num{0.15}\\
            Al-7075  & remainder & \num{0.4} & \num{0.5} & \numrange{1.2}{2.0} & \num{0.3} & \numrange{2.1}{2.9} & \numrange{0.18}{0.28} & \numrange{5.1}{6.1} & \num{0.2} & \num{0.15}\\
        
        \end{tabular}
    \end{center}
\end{table}

\subsection{Diagnostic Setup}\label{sec:diag}
\begin{figure}[ht!]
	\centering
	\includegraphics[width=.6\textwidth]{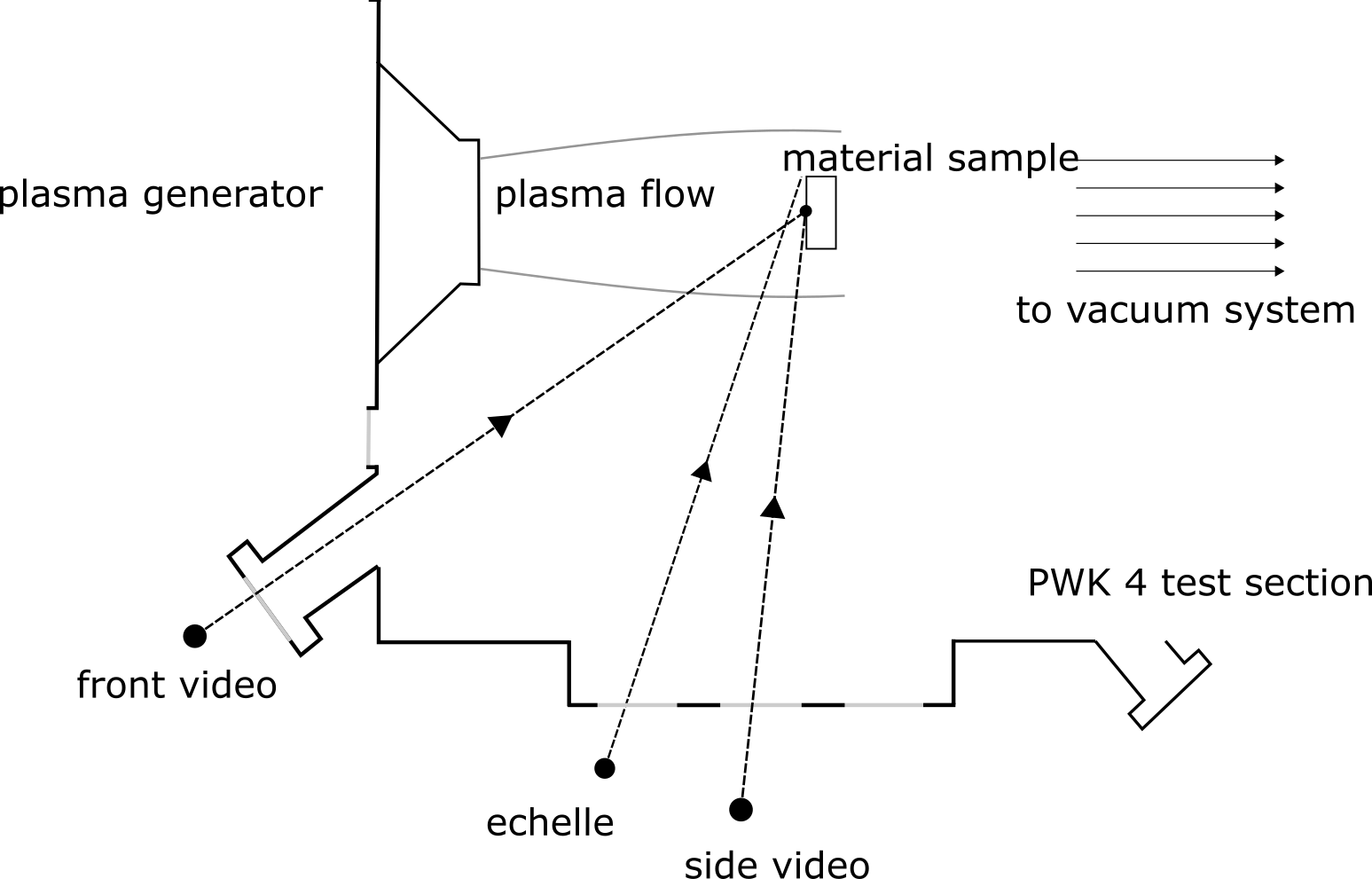}
	\caption{Schematic diagnostic setup of PWK4 (top view).}
	\label{fig:diag_setup}
\end{figure}

Figure~\ref{fig:diag_setup} shows the schematic of the experimental setup in PWK4 at IRS. The mechanical cylinder~\cite{Leiser_2019_01} holds the sample at the reference distance from the generator nozzle (see table~\ref{tab:generator_cond}). 

The front and side windows are used for \emph{Canon EOS 5DSR} digital cameras for a direct visual investigation. 
\begin{figure}[!ht]
	\centering
	\subfloat[Front view]{\includegraphics[trim= 30cm 0cm 30cm 0cm,clip,height = 5cm]{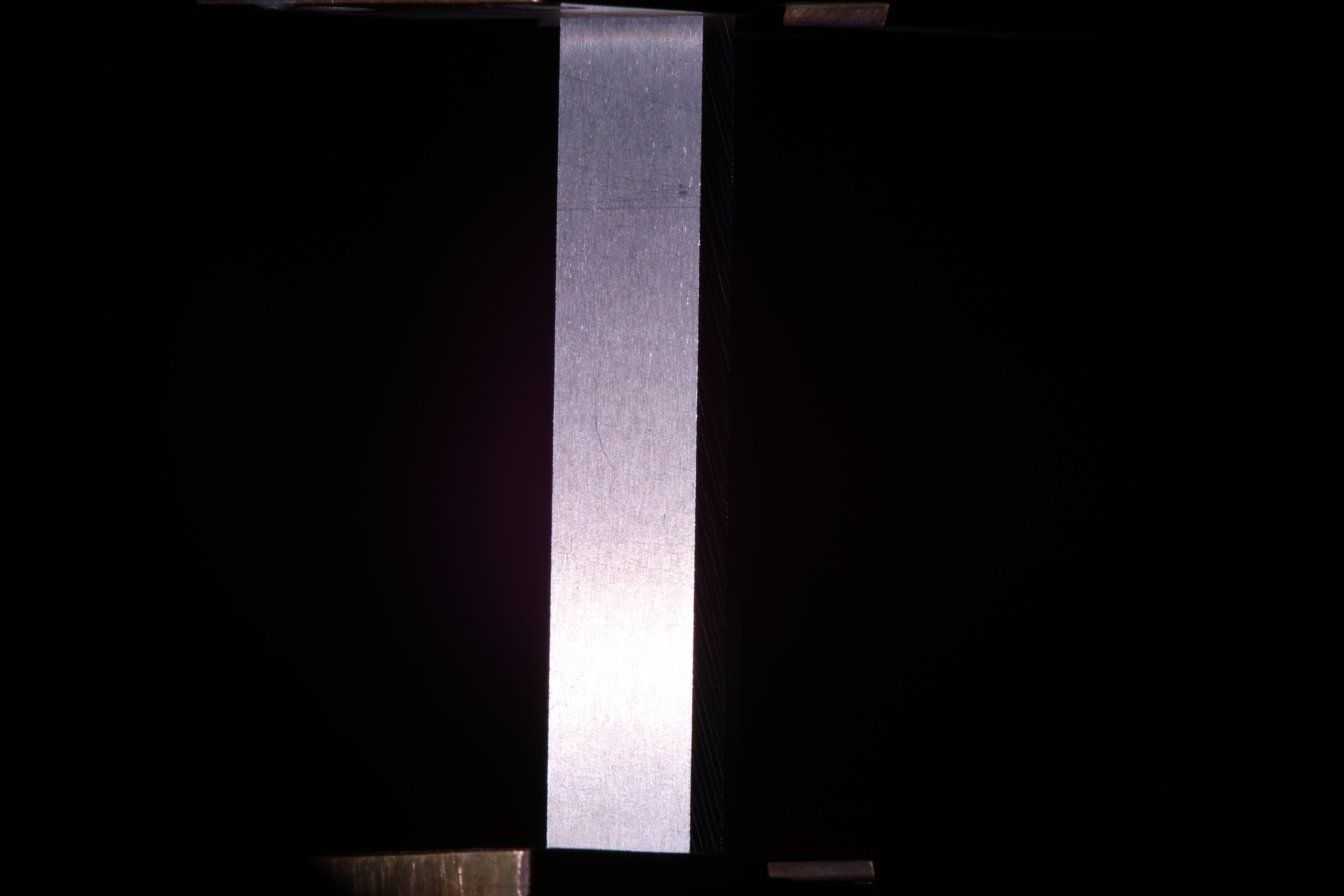}\label{fig:video_front}}\quad
	\subfloat[Side View]{\includegraphics[height = 5cm]{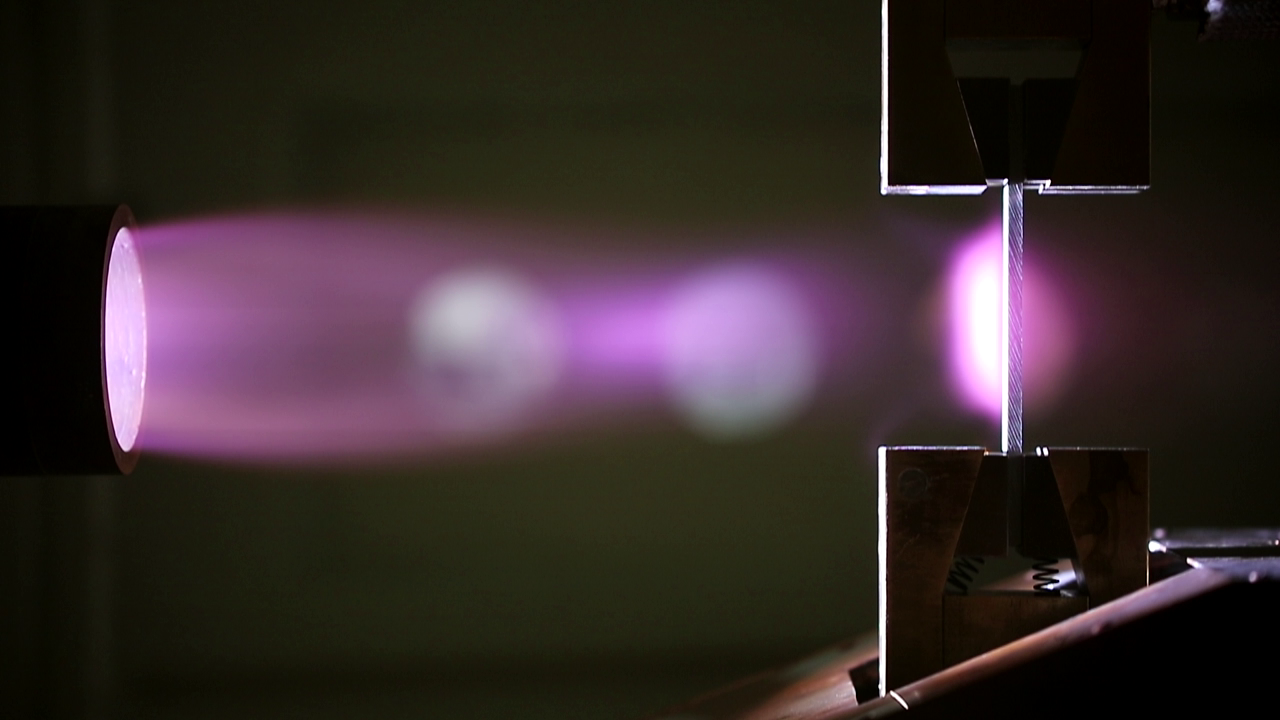}\label{fig:video_side}} 
	\caption{Still frame of the video imaging system \label{fig:video}}
\end{figure}
Sample images recorded by the video imaging system are shown in Fig.~\ref{fig:video}. The front view (Fig.~\ref{fig:video_front}) allows an in-depth analysis of the sample while the side view (Fig.~\ref{fig:video_side}) allows better insight into the plasma flow specifically discoloration.

\begin{figure}[ht!]
	\centering
	\includegraphics[width=.7\textwidth]{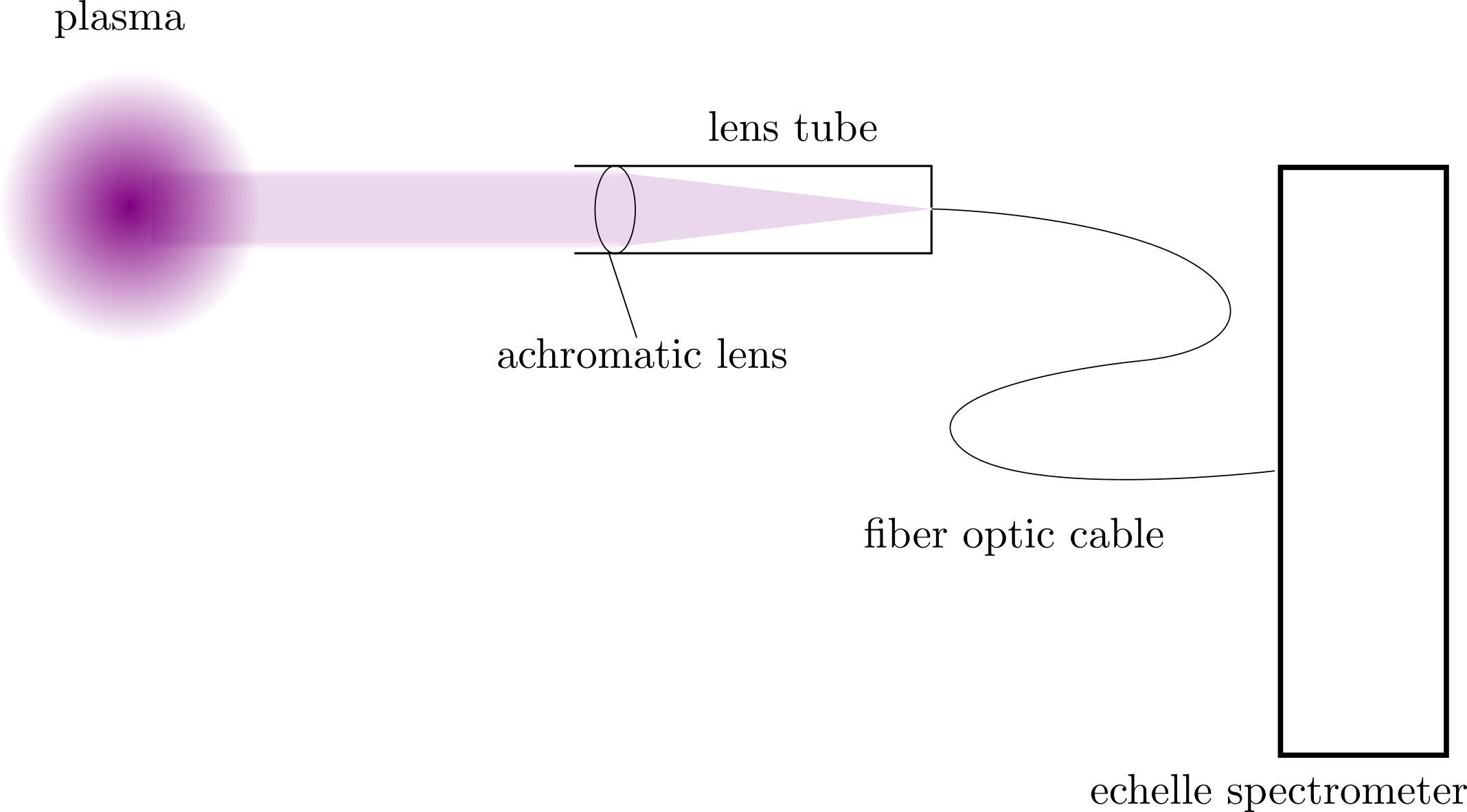}
	\caption{Overview of the optical setup of the Echelle spectrometer system.}
	\label{fig:diag_ECH}
\end{figure}

An \emph{LTB Aryelle 150} Echelle spectrometer~\cite{Loehle_2016_02} is used to record the spectra in the range of \SIrange{250}{880}{\nano\meter}. The setup of the optical system is shown in Fig.~\ref{fig:diag_ECH}. 

A \SI{50}{\milli\meter} lens tube with an achromatic lens optimized between \SIrange{450}{900}{\nano\meter} captures the emission radiation and focuses it onto a \SI{50}{\micro\meter} fiber optic cable port. This leads to the slitless spectrometer entry port. No further apertures or filters were used for this campaign.
%The spectrometer is aligned with an optical setup that focuses the line of sight into a \SI{50}{\micro\meter} fiber optic cable. This leads to the slitless spectrometer entry port. The optics consist of a \SI{50}{\milli\meter} lens tube with a quartz focussing lens. No further apertures or filters were used for this campaign.

The spectrometer optics are aligned to record the emissions from the stagnation point ahead of the surface of the material sample. This alignment prevents the spectrometer from recording the Planck radiation of the hot sample while capturing all of the immediate stagnation point emission in the gas phase. %While the spectrometer 

\begin{figure}[!ht]
	\centering
	\subfloat[Calibration lamp output curve]{\includegraphics[width = 0.492\textwidth]{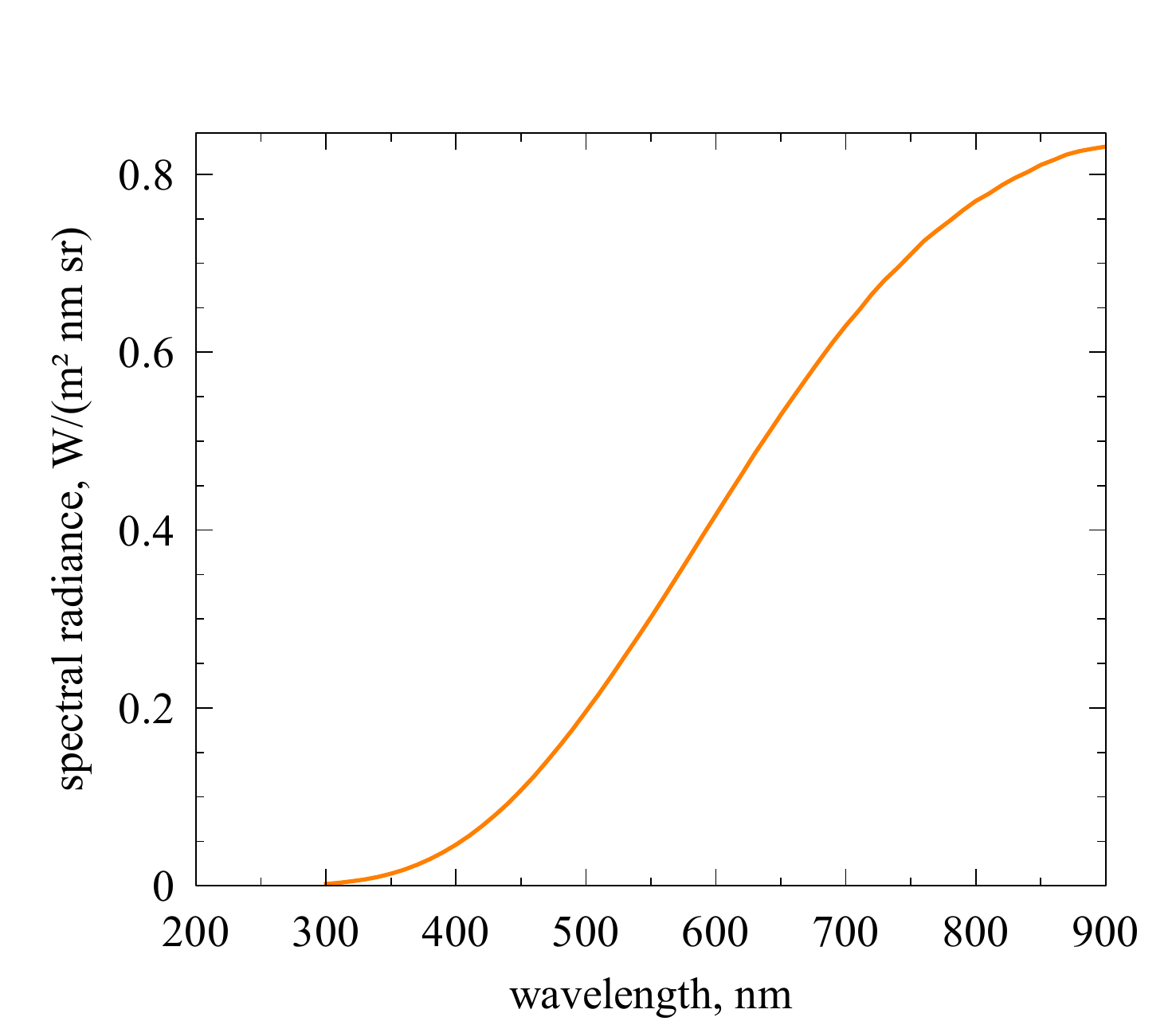}\label{fig:ECH_lamp}}
	\subfloat[Calibration spectra]{\includegraphics[width = 0.492\textwidth]{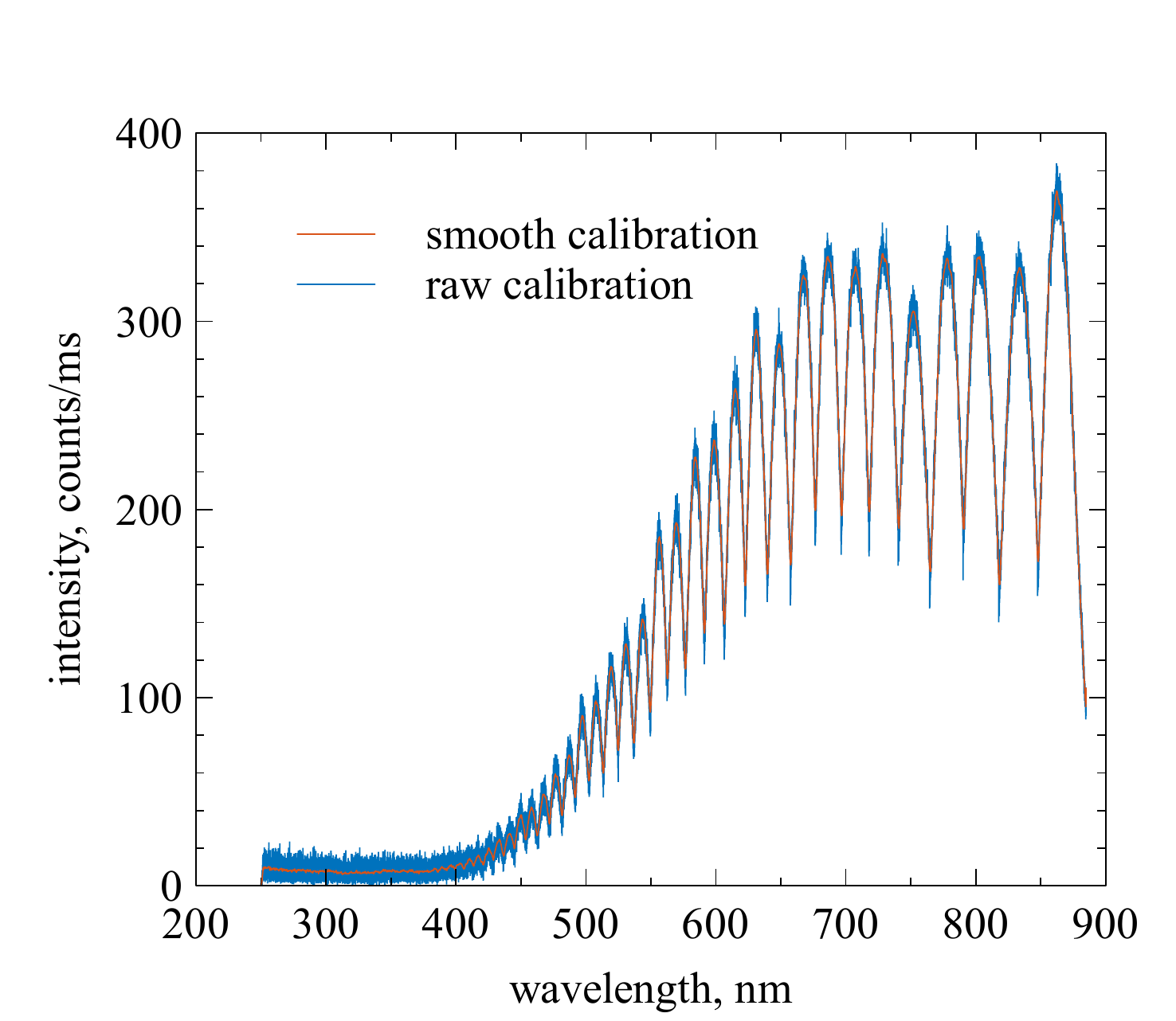}\label{fig:ECH_cal}}\\
	\subfloat[Time normalized spectrum]{\includegraphics[page=1,width = 0.492\textwidth]{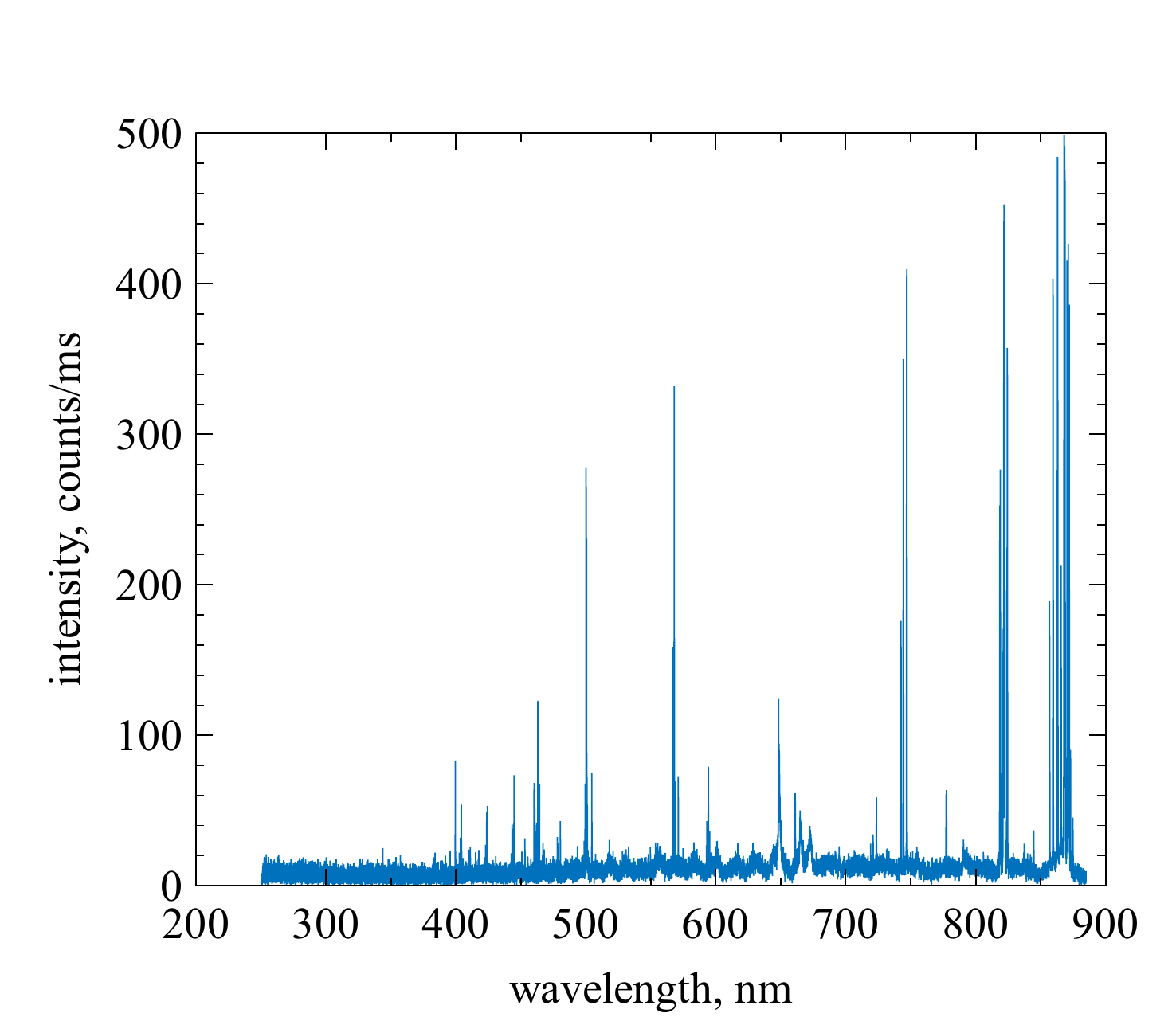}\label{fig:ECH_raw}} 
	\subfloat[Calibrated spectrum]{\includegraphics[page=2,width = 0.492\textwidth]{images/210913_Raw_cal.pdf}\label{fig:ECH_rawcal}} 
	\caption{Echelle calibration overview\label{fig:Echelle}}
\end{figure}

The advantage of an Echelle spectrometer is its high spectral resolution over a long wavelength interval. This is realized by taking spectra at very high orders (40-60) which subsequently are aligned vertically on the detector chip. The system used here was designed to detect UV-NIR without any gaps. Since it is used in airborne observations, a compact system is required. The camera connected to the spectrometer is an EMCCD (Qimaging Rolera) with a minimum exposure time of 1\,ms. Within one order the sensitivity has a central peak, so that aligning the spectra results in the spectra as shown in Fig.~\ref{fig:Echelle}. 

The wavelength of the spectrometer was calibrated with a mercury-argon lamp. The intensity calibration of the system is realized by positioning an Ulbricht sphere (Gigahertz Optics BN0102) at the location of interest in the tunnel. A spectrum is recorded (Fig.~\ref{fig:ECH_cal}) and time normalized. The known radiance of the calibration lamp (Fig.~\ref{fig:ECH_lamp}) results in a wavelength dependent calibration factor. This calibration factor can be applied to a time normalized spectrum  (Fig.~\ref{fig:ECH_raw}), resulting in a calibrated spectrum (Fig.~\ref{fig:ECH_rawcal}).
The sensitivity peaks within one spectral order can not be removed completely without loosing spectral information at the same time. Therefore the resulting spectra always contain a small fraction of this intensity distribution over the observed order. 

The manner in which the diffraction orders are projected onto the camera chip can also lead to order cross-talk, meaning that the software attributes the signal to the wrong diffraction order and therefore the wrong wavelength. Such an error occurred during parts of this campaign. The effect appeared at wavelengths upwards of \SI{550}{\nano\meter}. While this makes the absolute intensity calibration impossible, the relative temporal evolution of lines is not affected by this error.

\begin{figure}[hbt!]
	\centering
	\includegraphics[width=.6\textwidth]{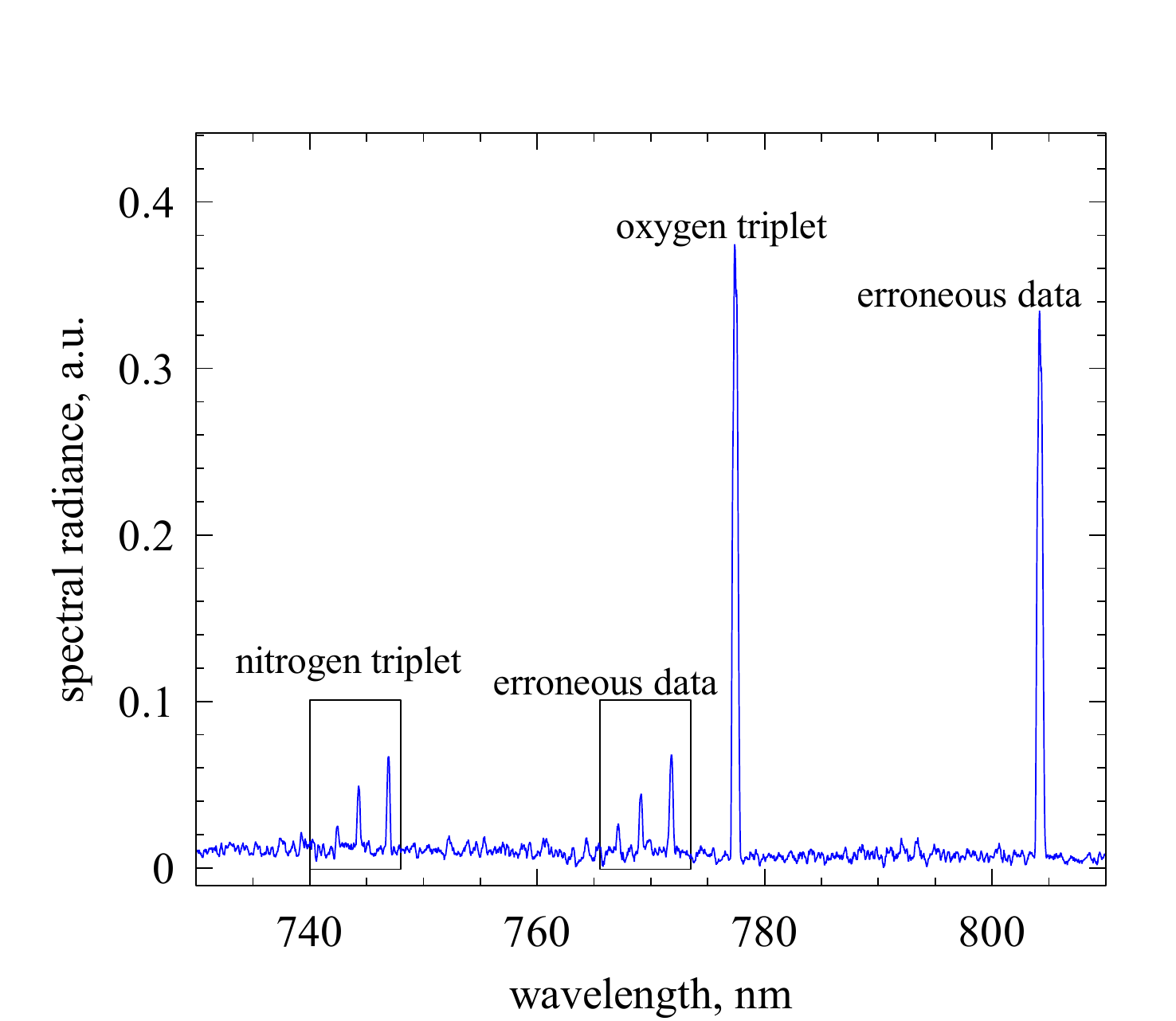}
	\caption{Sample spectrum.}
	\label{fig:sample_spec}
\end{figure}

A sample spectrum of a stagnation point plasma on a cool wall is shown in Fig.~\ref{fig:sample_spec} in the wavelength range of \SIrange{730}{810}{\nano\meter} to show the prominent atomic lines that correspond to oxygen triplet at \SI{777}{\nano\meter} and the nitrogen lines \SIrange{742}{747}{\nano\meter}.%between \SI{700}{\nano\meter} and \SI{800}{\nano\meter}. 

The misinterpreted lines are visible at \SI{804}{\nano\meter}, which can be attributed to the \SI{777}{\nano\meter} oxygen triplet as well as erroneous lines between \SIrange{765}{775}{\nano\meter}, which correspond to the nitrogen lines between \SIrange{742}{747}{\nano\meter}. Spectra that show this double appearance of lines are annotated in section~\ref{sec:results}. 

After the error was identified, the spectrometer system was readjusted. This led to a better signal, which allowed the system to be operated with a lower gain than during the initial part of the campaign. The lower gain led to less pronounced noise in the data and a better signal to noise ratio at comparable exposure times in the latter half of the campaign.

\section{Spectral Data Analysis}\label{sec:results}

The temporal development of the line intensity was determined by integrating the spectral lines over its total width. % the FWHM. 
For this, the central line position was determined around the expected spectral position (see table~\ref{tab:lambda_data}) taken from the NIST database~\cite{NIST_2013}. 

\begin{table}[ht]
	\begin{center}
		\caption{Center wavelength of the evaluated spectral lines for each species~\cite{NIST_2013}.}\label{tab:lambda_data}\vspace{1em}
		\sisetup{table-number-alignment = center,table-figures-integer  = 1,table-figures-decimal  = 1}
		\renewcommand{\arraystretch}{1.2}
		\begin{tabular}{ccSc}
		element & symbol & {wavelength, \si{\nano\meter}} & transition\\
		\midrule
		oxygen      & \ce{O} & 777.42  & (triplet) $3p-3s$\\
		nitrogen    & \ce{N} & 746.90 & $3p-3s$\\
		aluminum   & \ce{Al} & 396.15 & $4s-3p$\\
		magnesium   & \ce{Mg} & 518.36 & $4s-3p$\\
		silicon     & \ce{Si} & 728.92 & $3p(^2P°_{3/2})4f-3p^3$\\
		zinc        & \ce{Zn} & 481.05 & $5s-4p$\\
		copper      & \ce{Cu} & 521.82 & $4d-4p$\\
		iron        & \ce{Fe} & 438.35 & $4p-4s$\\
		lithium     & \ce{Li} & 670.78 & (doublet) $2p-2s$\\
		sodium      & \ce{Na} & 589.00 & $3p-3s$\\
		potassium   & \ce{K} & 766.49 & $4p-4s$\\
		
\end{tabular}
\end{center}
\end{table}

\begin{figure}[ht!]
	\centering
	\includegraphics[width=.6\textwidth]{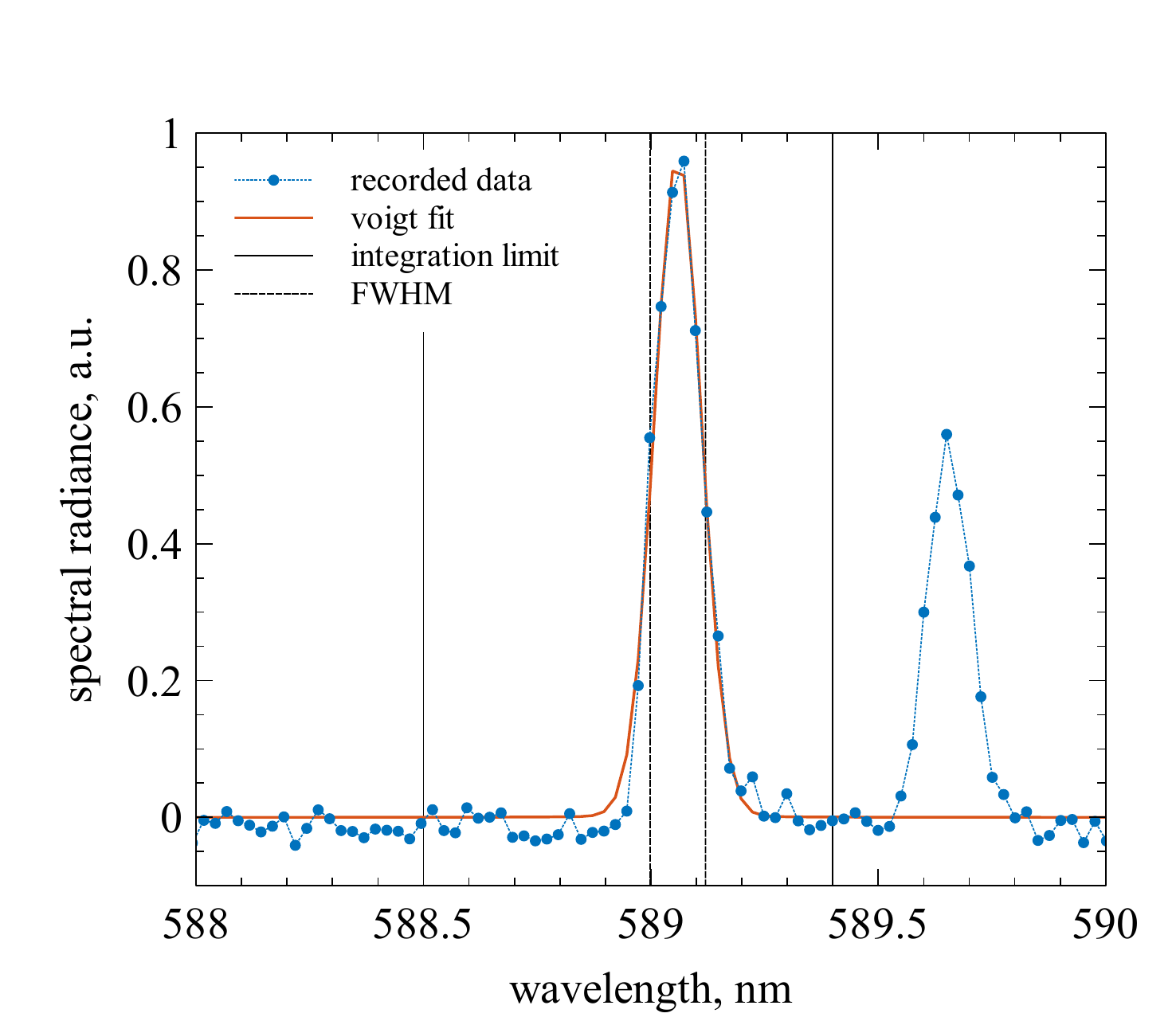}
	\caption{Recorded data with a Voigt fit of a sodium line, FWHM and integration limits.}
	\label{fig:voigt}
\end{figure}

The line width was determined by fitting a Voigt function for the detected lines. The Voigt function is approximated using the methods of Liu~et.~al~\cite{Liu_2001_01} as well as Olivero and Longbothum~\cite{Olivero_1977_01}. Using a non-linear regression algorithm the Gaussian and Lorenzian widths are determined for each spectrum.
The result of a line determination and fit is shown exemplary at a sodium line in Fig.~\ref{fig:voigt}. The integration limits were chosen to avoid other lines close to the observed one.

Molecular bands were identified using the database by Pearse and Gaydon~\cite{pearse}, these were not evaluated further.
%In case of the oxygen line, this is in fact a triplet that overlaps in the recorded data, the singular lines are not distinguishable. 
%The qualitative development of the line strength was determined by integrating the spectral lines at the expected position. The width of each line was not determined, the lines were integrated over the center wavelength \SI{+-0.5}{\nano\meter}. The center wavelength of the lines that were evaluated displayed in tab.~\ref{tab:lambda_data} for each element. These were taken from the NIST database~\cite{NIST_2013}. 

The onset of the experiment and test time $t = \SI{0}{\second}$ is determined by the probe reaching the center of the flow.
The atomic lines and bands that are visible in all experiments correspond to \ce{N}, \ce{N+}, \ce{N2}, \ce{N2+}, \ce{O}, and \ce{O2}, as expected from the air plasma. The variations of intensity over time are weak and common for such experiments~\cite{Hermann_2016_04}.

\subsection{Aluminum 6060}

\subsubsection{65km}
\begin{figure}[!ht]
	\centering
	\subfloat[Front view]{\includegraphics[height = 7cm]{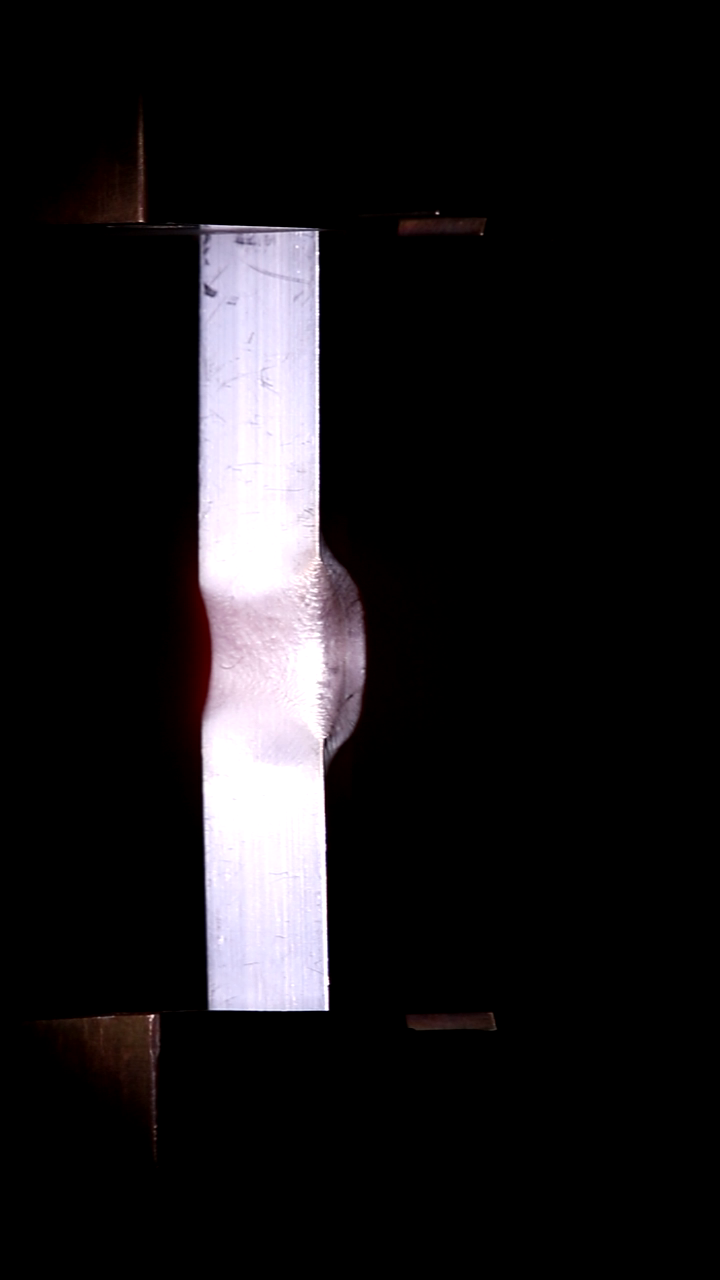}\label{fig:AL6_front}}\quad
	\subfloat[Side View]{\includegraphics[trim= 0cm 0cm 8cm 0cm,clip,height = 7cm]{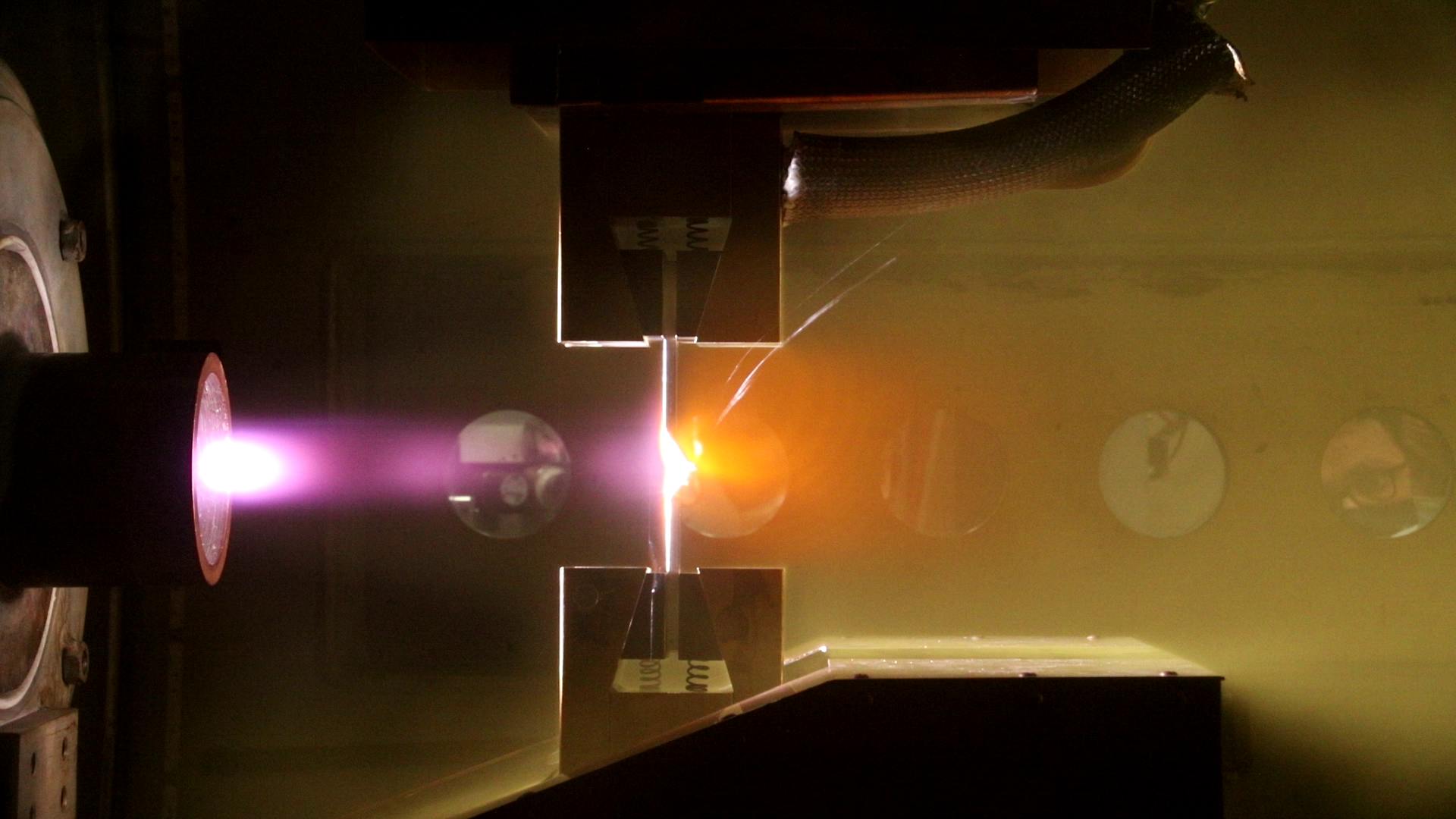}\label{fig:AL6_side}} 
	\caption{Still frame of aluminum 6060 test at 65\,km Scenario 1 \label{fig:AL6_vid}}
\end{figure}

Figure~\ref{fig:AL6_vid} shows a still frame of the aluminum 6060 test at \SI{65}{\kilo\meter} with force scenario 1 shortly before the material failure. The front view (Fig.~\ref{fig:AL6_front}) shows the  bulk material melting with the thin oxide layer holding the sample together, the deformation of the sample surface is visible here. An image shortly later is shown in the side view (Fig.~\ref{fig:AL6_side}), with molten material being ejected into the wake. While a strong discoloration is especially visible in the wake, it is also visible in the stagnation region. 

\begin{figure}[!h]
	\centering
	\subfloat[Scenario 1 spectra at five test times of interest, marked in Fig.~\ref{fig:Al6_fl_65_Sz1_ECH_time}]{\includegraphics[width = 0.492\linewidth]{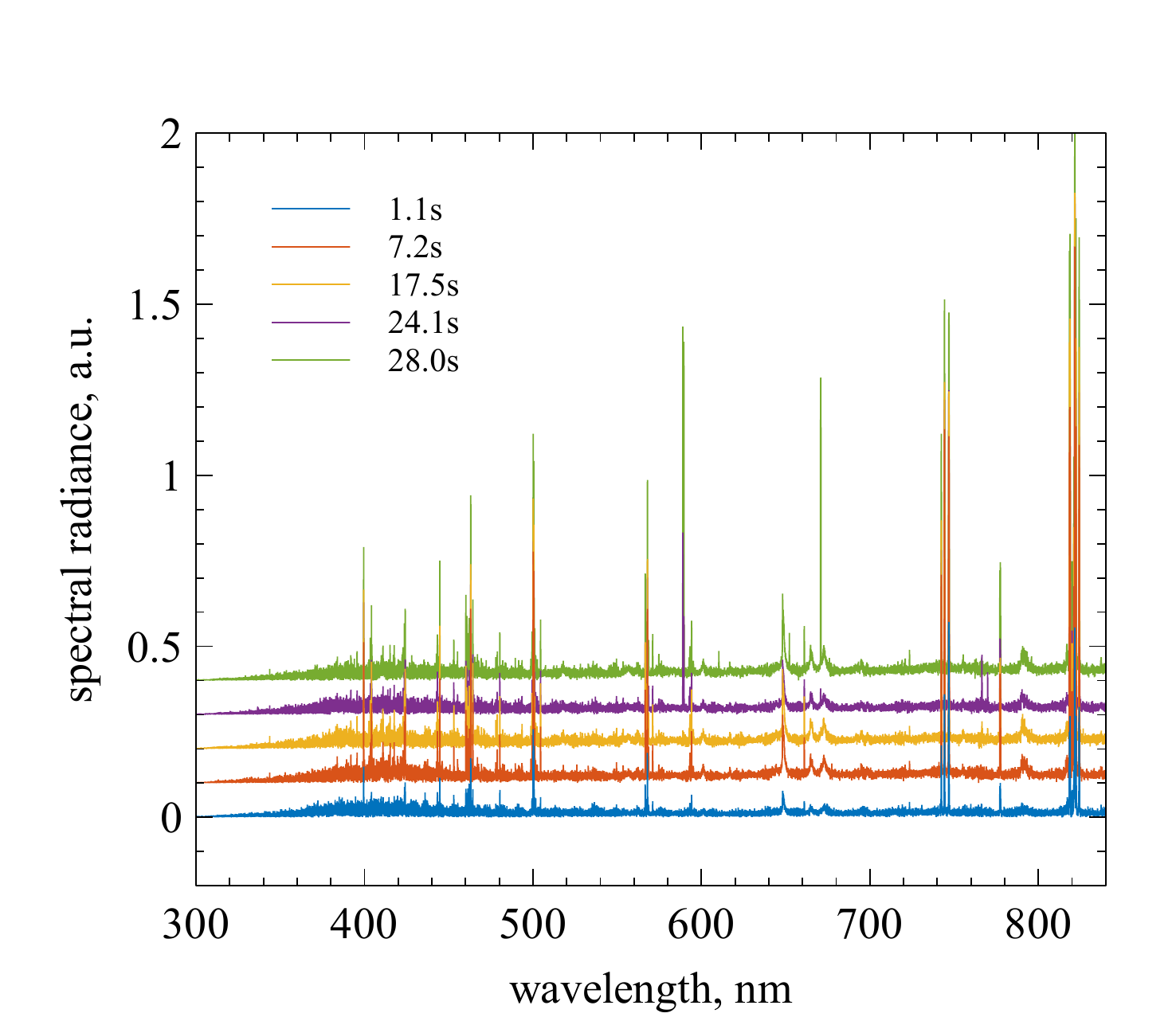}\label{fig:Al6_fl_65_Sz1_ECH_spectra}}
	\subfloat[Scenario 1 spectral radiance of the individual lines over time]{\includegraphics[width = 0.508\linewidth]{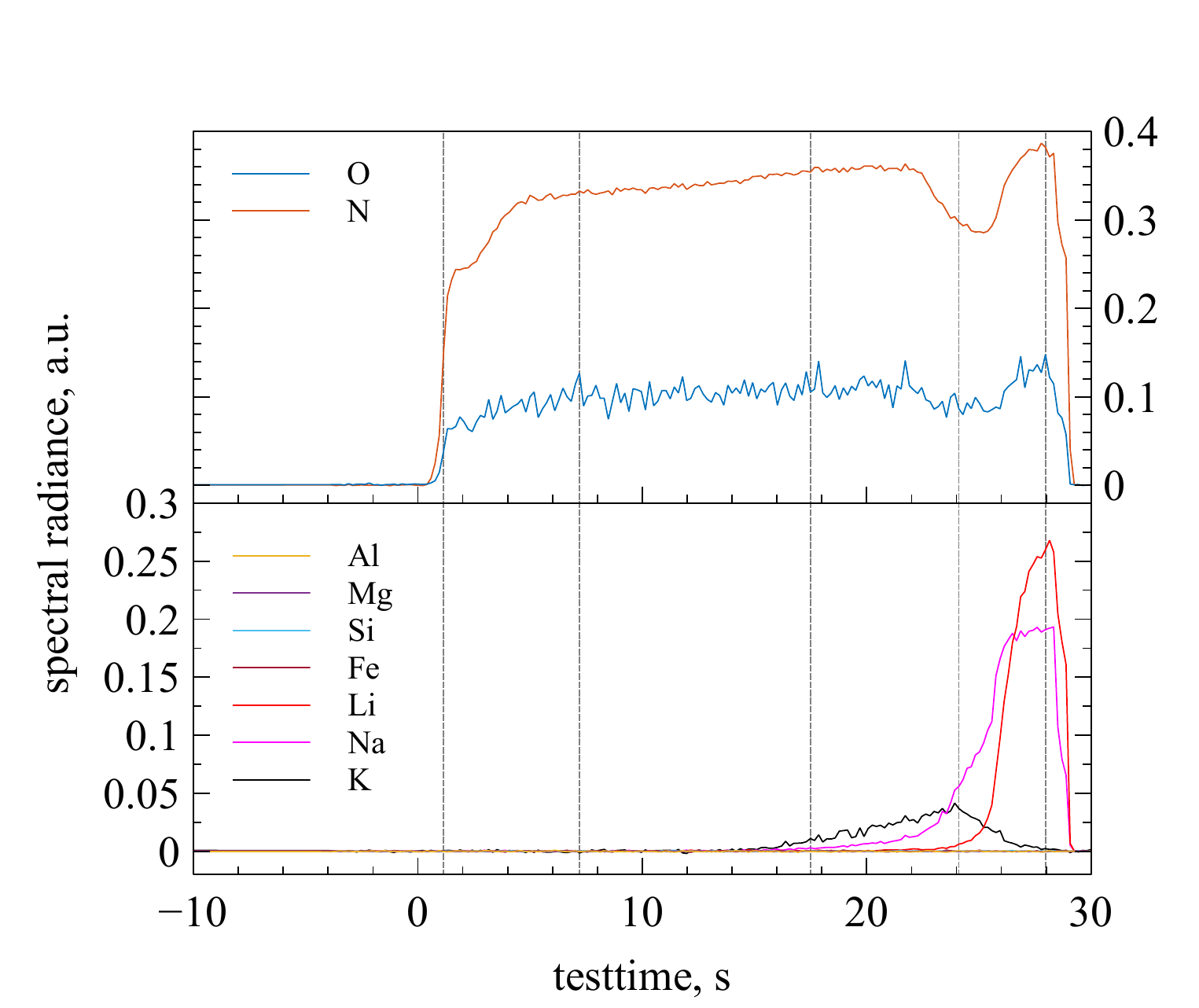}\label{fig:Al6_fl_65_Sz1_ECH_time}} \\
	\subfloat[Scenario 2 spectra at five test times of interest, marked in Fig.~\ref{fig:Al6_fl_65_Sz2_ECH_time}]{\includegraphics[width = 0.492\linewidth]{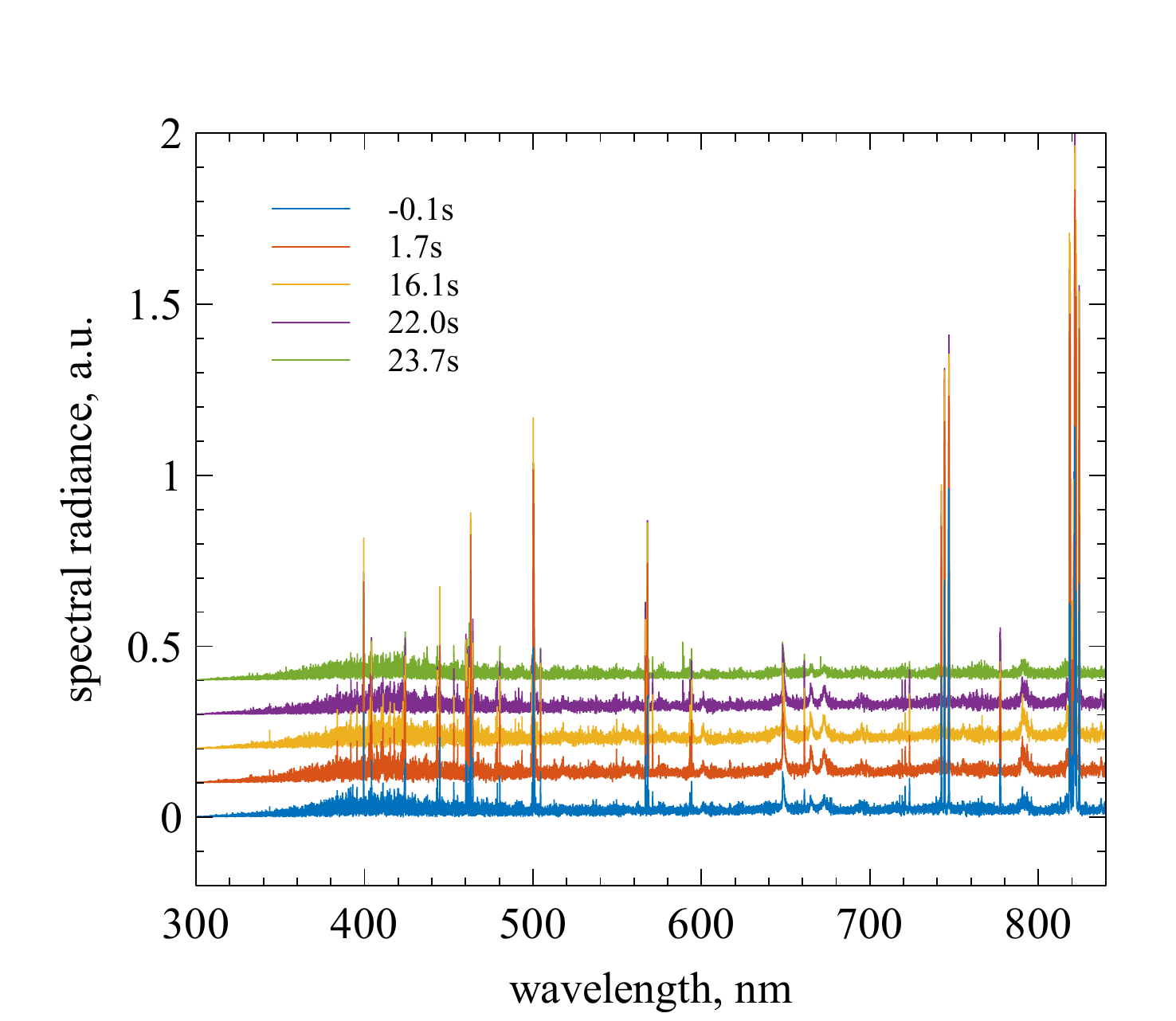}\label{fig:Al6_fl_65_Sz2_ECH_spectra}}
	\subfloat[Scenario 2 spectral radiance of the individual lines over time]{\includegraphics[width = 0.508\linewidth]{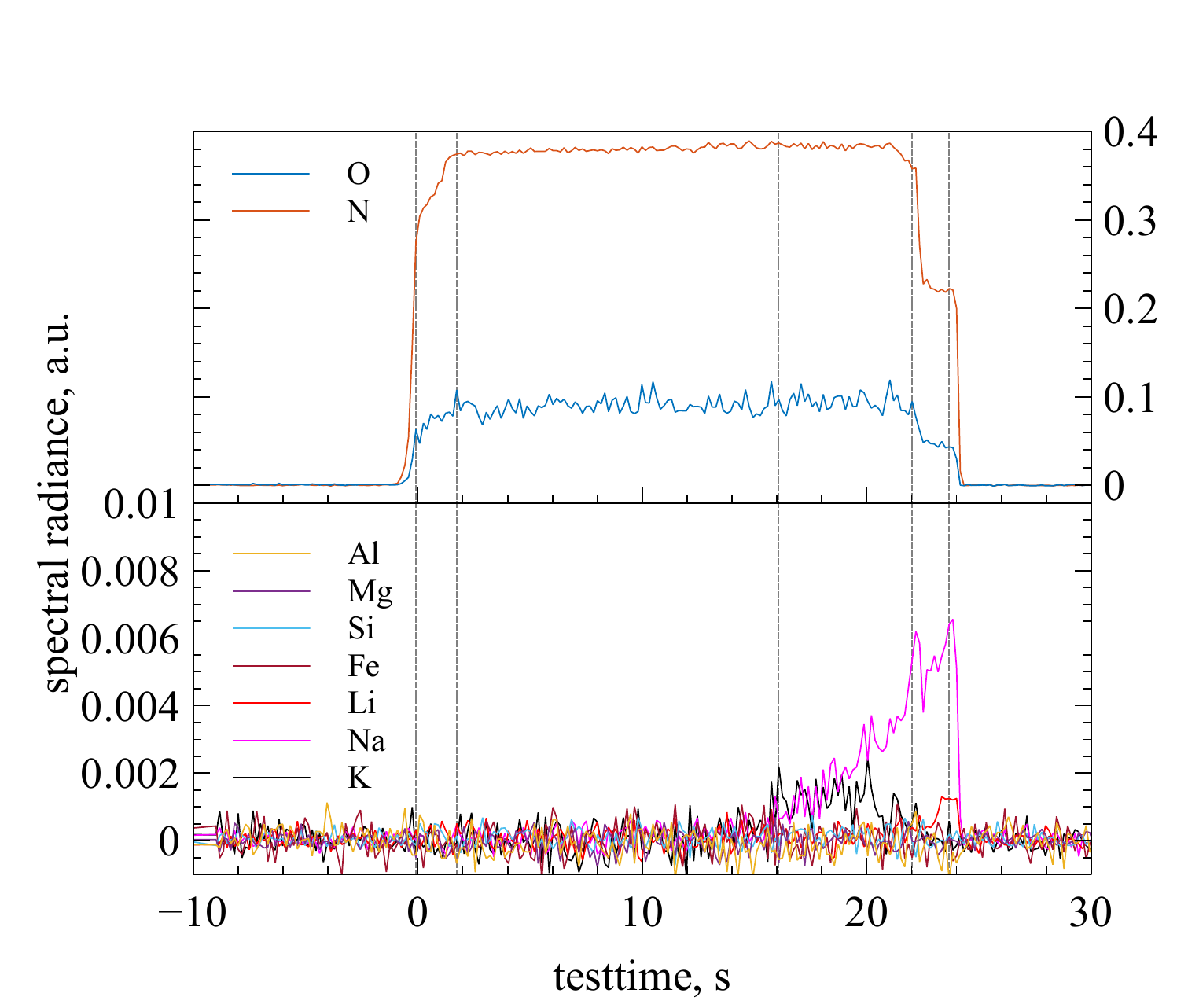}\label{fig:Al6_fl_65_Sz2_ECH_time}} 
	\caption{Echelle data of Al-6060 at 65km\label{fig:Al6_fl_65_ECH}}
\end{figure}

Spectral data collected during the \SI{65}{\kilo\meter} experiments is shown in Fig.~\ref{fig:Al6_fl_65_ECH}. Figure \ref{fig:Al6_fl_65_Sz1_ECH_spectra} shows five sample spectra recorded throughout the experiment conducted with negligible force (scenario 1). For better visibility, the spectra are offset to one another. The temporal evolution of the integrated line intensity is shown in Fig.~\ref{fig:Al6_fl_65_Sz1_ECH_time} with the air plasma lines shown on top and the alloy constituents on the bottom. Spectra recorded during the experiment with nominal forces (scenario 2) applied is displayed in Fig.~\ref{fig:Al6_fl_65_Sz2_ECH_spectra}, the respective time evolution is shown in Fig.~\ref{fig:Al6_fl_65_Sz2_ECH_time}. %The low framerate in the first \SI{12}{\second} of the experiment is due to an adaptation of the exposure time to the test condition. 

During both tests, no spectral signatures of the alloy were visible. A weak emission line corresponding to \ce{K} appears, after around \SI{15}{\second} in both experiments. Shortly afterwards the prominent \ce{Na} line appears and gradually gets stronger. In scenario 2 the probe fails due to the applied forces, at around \SI{23}{\second} while in scenario 1 the \ce{Na} line intensity increases until it saturates the spectrometer at \SI{26}{\second}. In the scenario 1 test, at around \SI{24}{\second}, the bulk material melted, determined by the video data which corresponds to the \ce{Li} line appearing. This line grows in intensity, until the material failed at the end of experiment, $t=\SI{28}{\second}$. In scenario 2, \ce{Li} is visible for a few frames very weakly shortly before the sample fails.
%24s in melting
%22s lithium
%15.8s sodium and potassium

\subsubsection{75km}
\begin{figure}[!ht]
	\centering
	\subfloat[Scenario 1 spectra at five test times of interest, marked in Fig.~\ref{fig:Al6_fl_75_Sz1_ECH_time}]{\includegraphics[width = 0.492\linewidth]{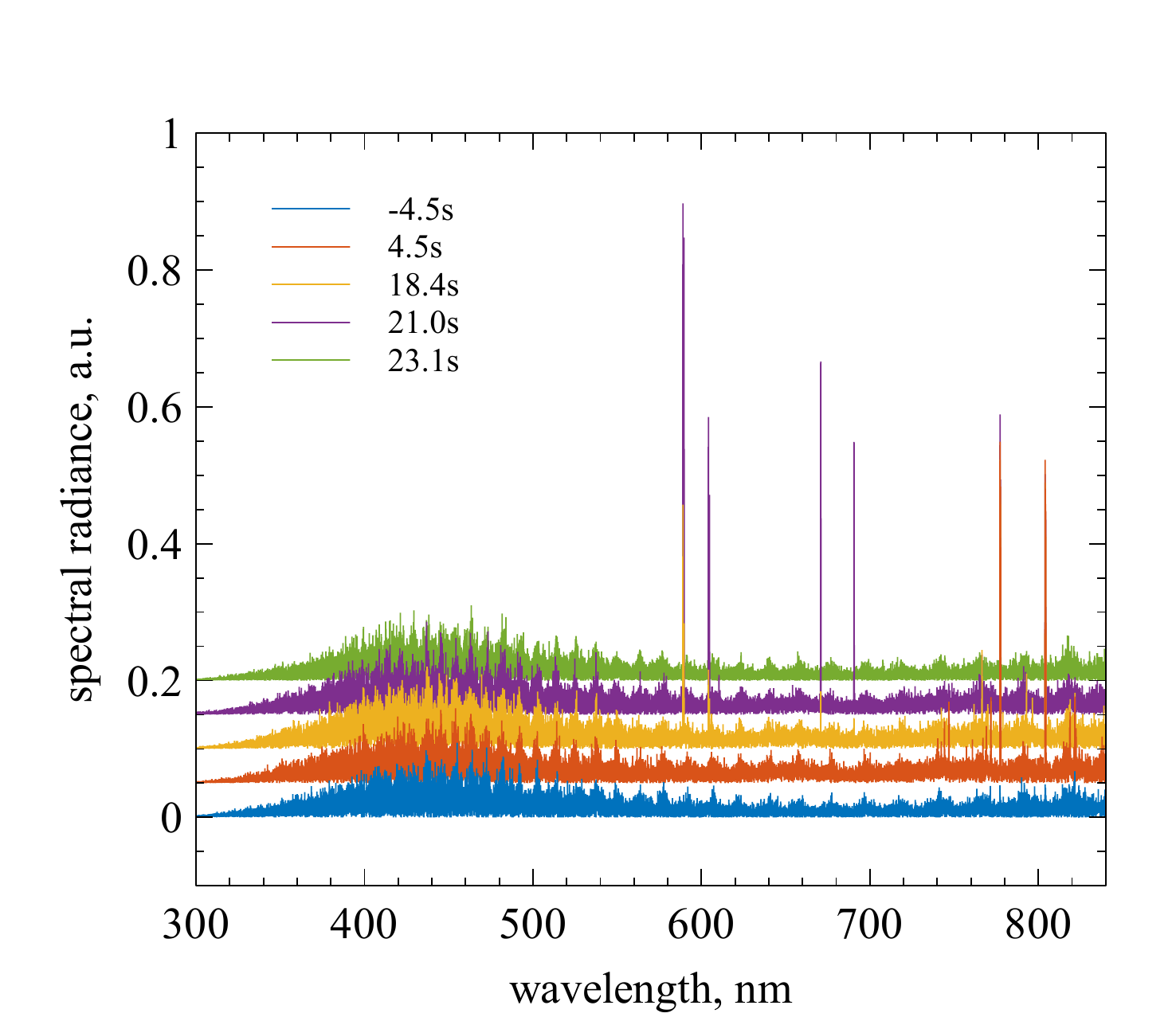}\label{fig:Al6_fl_75_Sz1_ECH_spectra}}
	\subfloat[Scenario 1 spectral radiance of the individual lines over time]{\includegraphics[width = 0.508\linewidth]{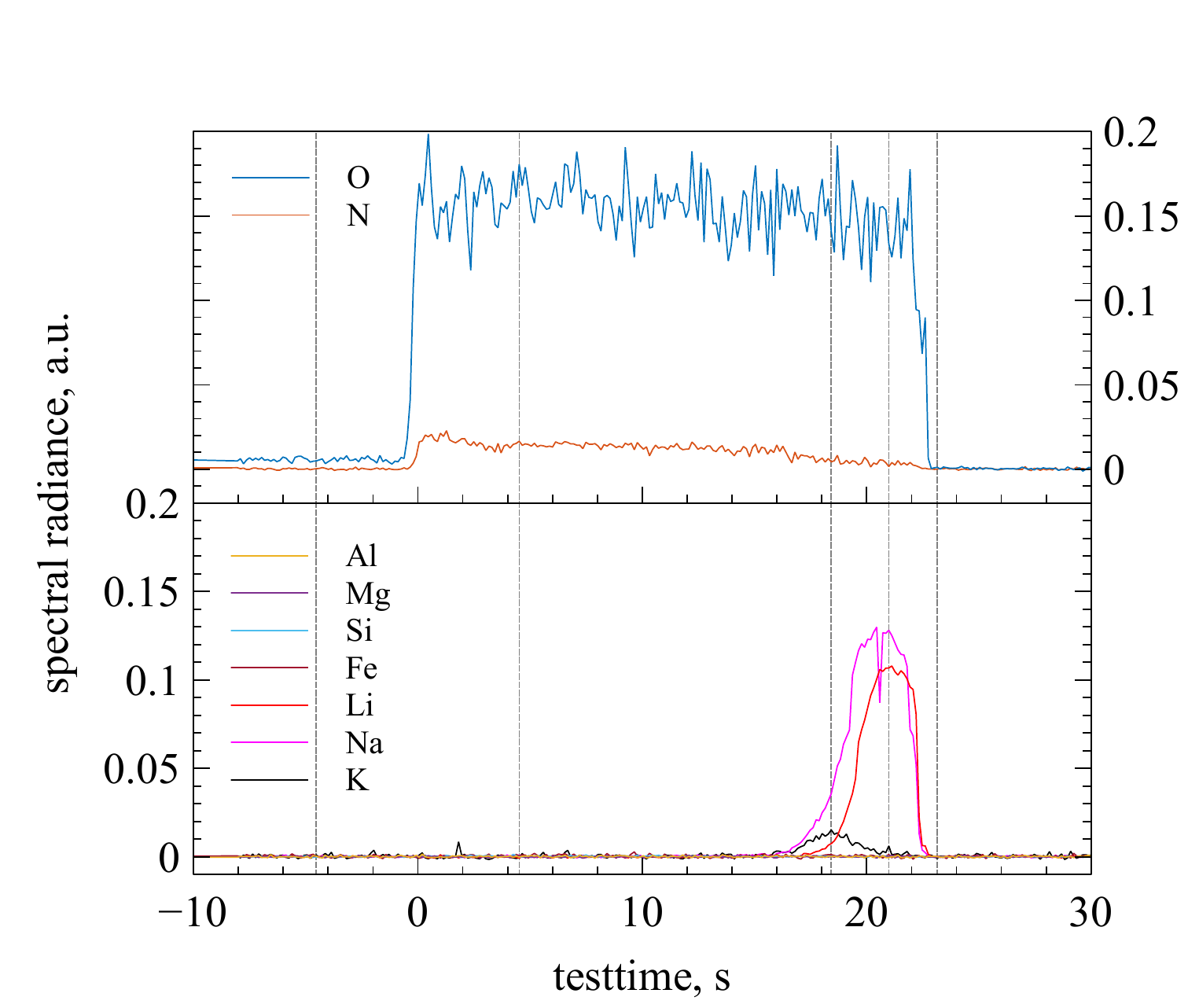}\label{fig:Al6_fl_75_Sz1_ECH_time}} \\
	\subfloat[Scenario 2 spectra at five test times of interest, marked in Fig.~\ref{fig:Al6_fl_75_Sz2_ECH_time}]{\includegraphics[width = 0.492\linewidth]{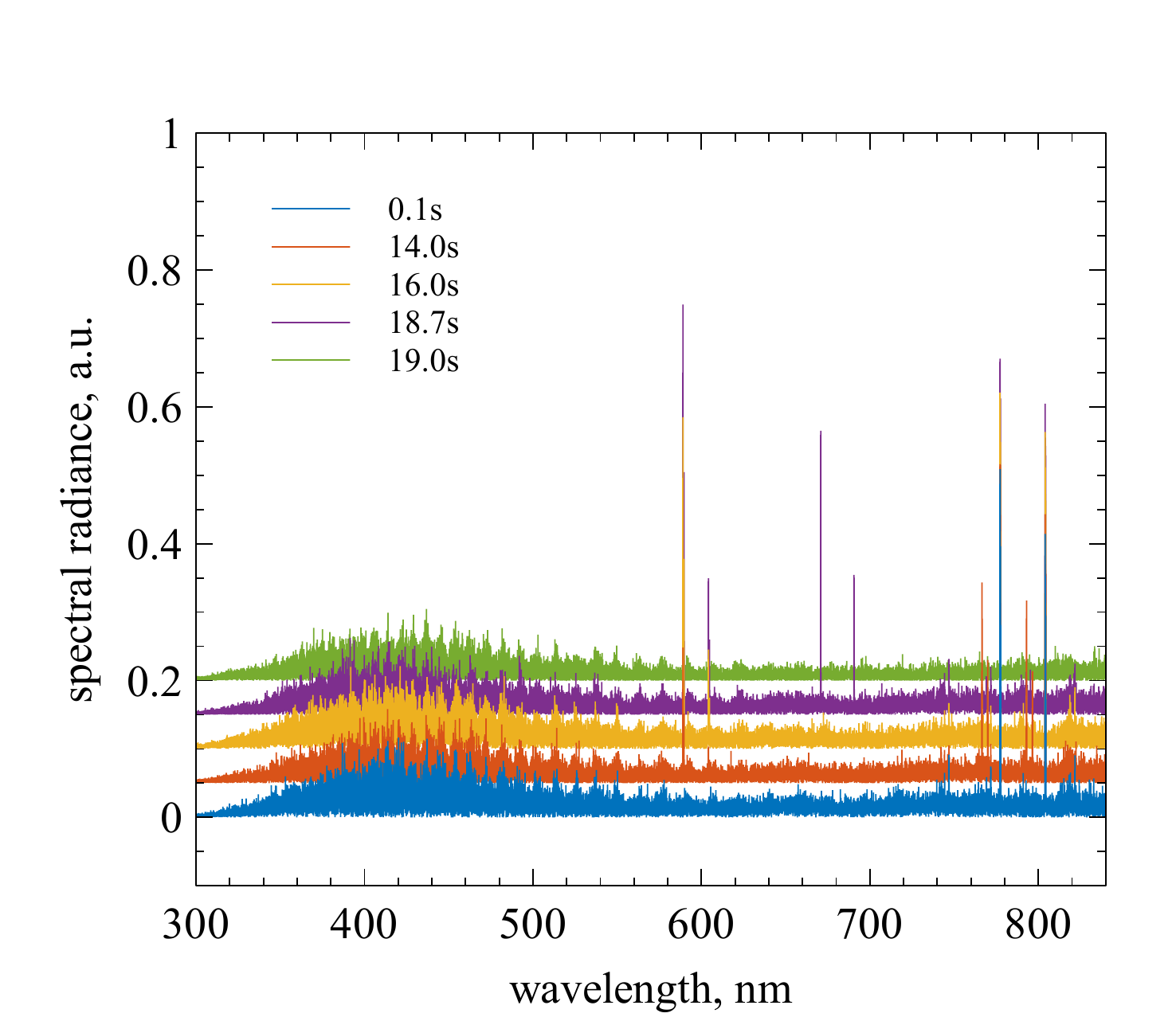}\label{fig:Al6_fl_75_Sz2_ECH_spectra}}
	\subfloat[Scenario 2 spectral radiance of the individual lines over time]{\includegraphics[width = 0.508\linewidth]{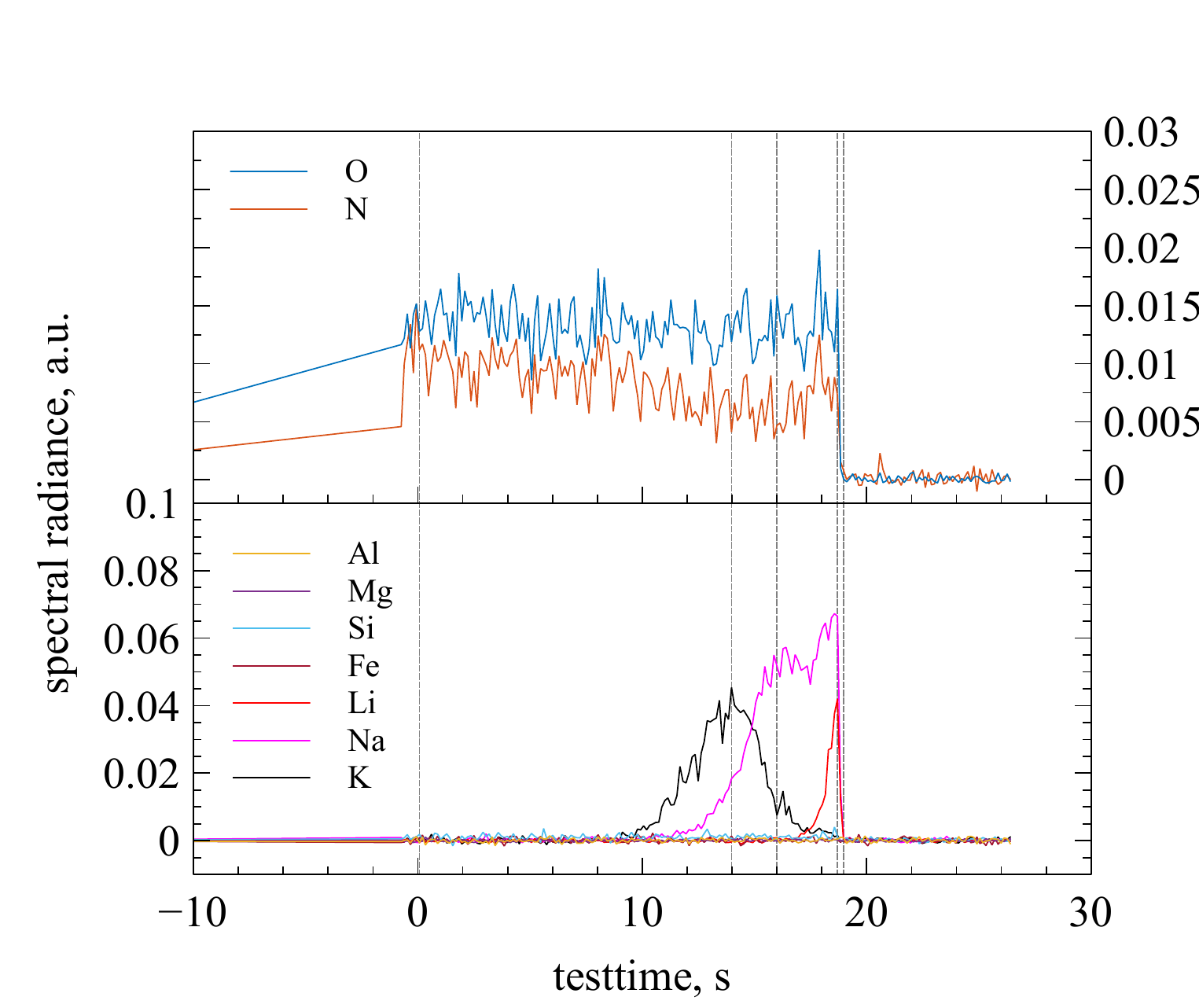}\label{fig:Al6_fl_75_Sz2_ECH_time}} 
	\caption{Echelle data of Al-6060 at 75km\label{fig:Al6_fl_75_ECH}}
\end{figure}

Data collected during the experiments at \SI{75}{\kilo\meter} is shown in Fig.~\ref{fig:Al6_fl_75_ECH}. These experiments were conducted during the first part of the campaign. 

Similar effects as during the \SI{65}{\kilo\meter} condition can be seen here, no alloy components are spectrally active, for the beginning of the test it is purely the plasma constituents. During the test a weak \ce{K} signature was detected, starting at \SI{15}{\second} during the scenario 1 test and at \SI{10}{\second} during the scenario 2 test. In both tests the intensity peaks within \SI{3}{\second} and diminishes over a further \SI{4}{\second}.
\ce{Na} is also discernible in both experiments with the onset fairly similar at \SI{16}{\second} and \SI{13}{\second} respectively.
During the scenario 1 test the intensity reaches a plateau between \SIrange{20}{21}{\second} and diminishes together with material deformation and failure. During scenario 2, the intensity of \ce{Na} increases towards the failure time, though the slope of the line seems to indicate that the peak intensity and a potential plateau lies around the failure time.
In both tests, \ce{Li} appears at around \SI{18}{\second}, with the intensity increasing rapidly. During scenario 1 the intensity peaks around \SI{21}{\second} and diminishes with the material deformation and failure, while during scenario 2 the probe fails with the intensity reaching a peak.

\subsubsection{90km}
\begin{figure}[!ht]
	\centering
	\subfloat[Spectra at five test times of interest, marked in Fig.~\ref{fig:Al6_fl_90_Sz2_ECH_time}]{\includegraphics[width = 0.492\linewidth]{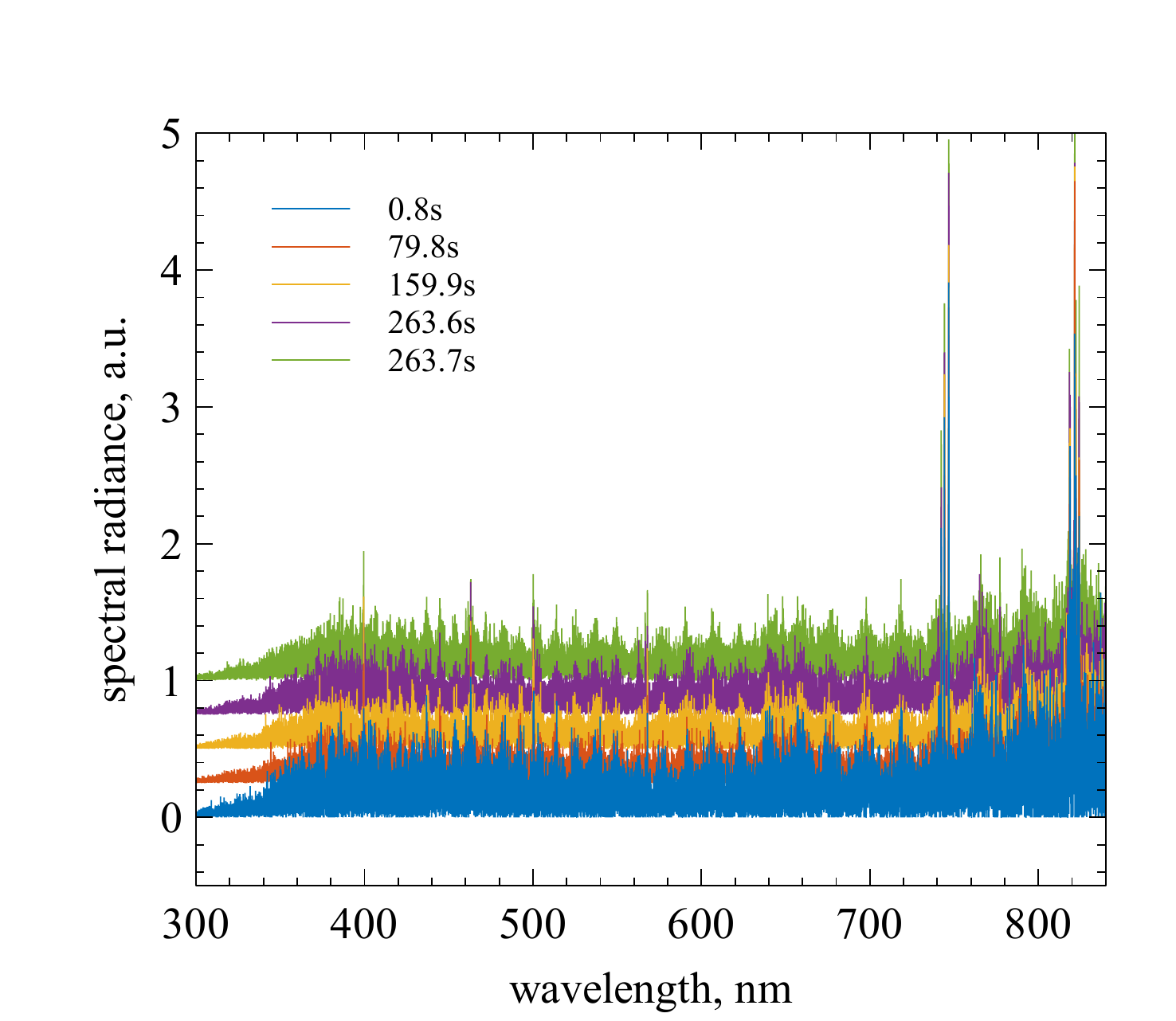}\label{fig:Al6_fl_90_Sz2_ECH_spectra}}
	\subfloat[Spectral radiance of the individual lines over time]{\includegraphics[width = 0.508\linewidth]{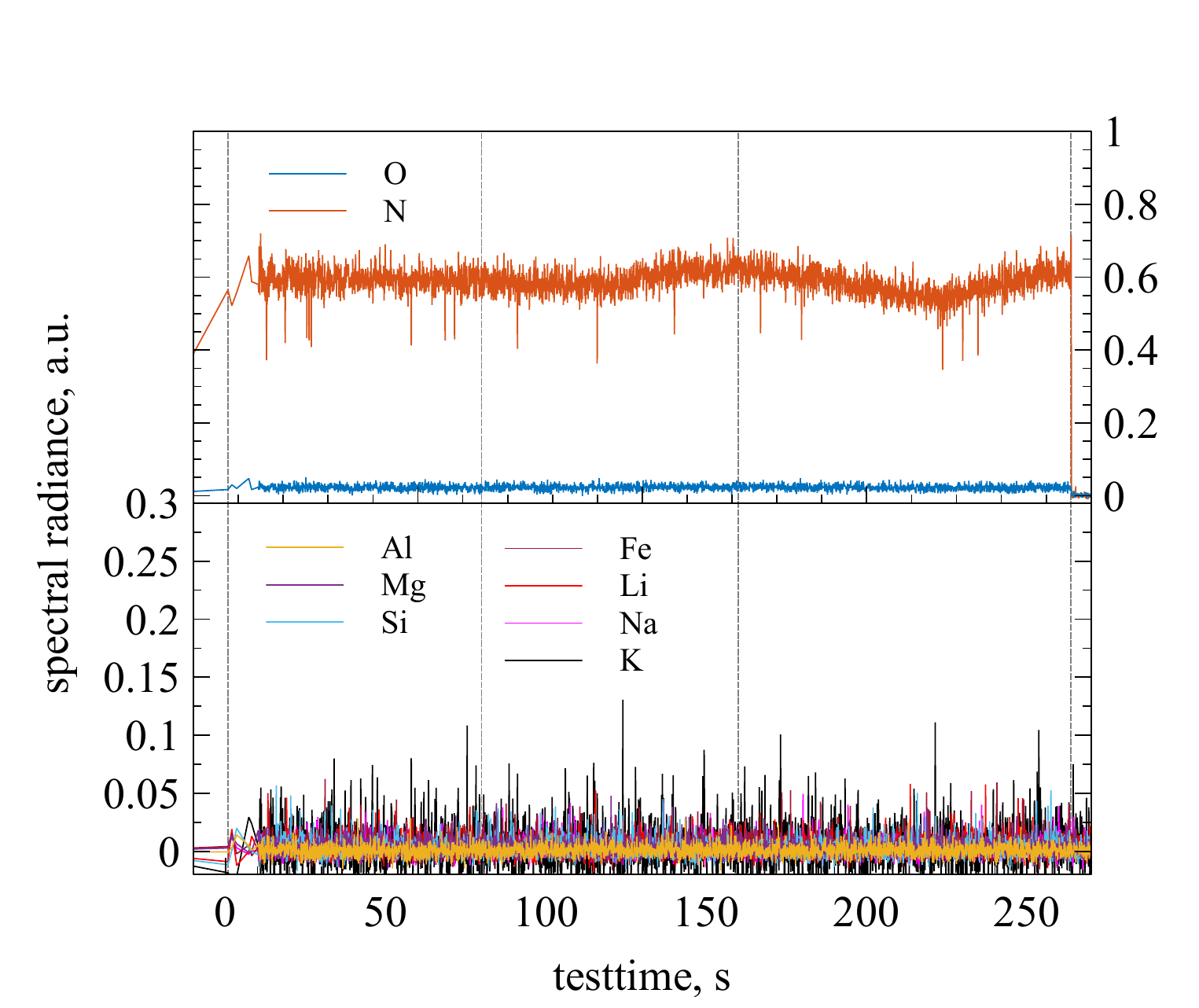}\label{fig:Al6_fl_90_Sz2_ECH_time}} 
	\caption{Echelle data of Al-6060 at 90km Sz1/2 \label{fig:Al6_fl_90_Sz2_ECH}}
\end{figure}

Data collected during the \SI{90}{\kilo\meter} condition experiment is shown in Fig.~\ref{fig:Al6_fl_90_Sz2_ECH}. Since the material sample did not fail at the nominal force (load scenario 1), the force was further increased until the material failed.%The experiment combined both the scenario 1 and 2 condition due to the sample not failing during scenario 2.
The data collected during this test is very noisy, due to the high gain settings chosen for the very weak emission. More importantly, the condition exhibits a low total pressure and low heat flux, which results in very weak plasma lines. Aside from these, no lines were distinguishable during testing.

\cleardoublepage
\subsection{Aluminum 7075}

\subsubsection{65km}

\begin{figure}[!ht]
	\centering
	\subfloat[Scenario 1 spectra at five test times of interest, marked in Fig.~\ref{fig:Al7_fl_65_Sz1_ECH_time}]{\includegraphics[width = 0.492\linewidth]{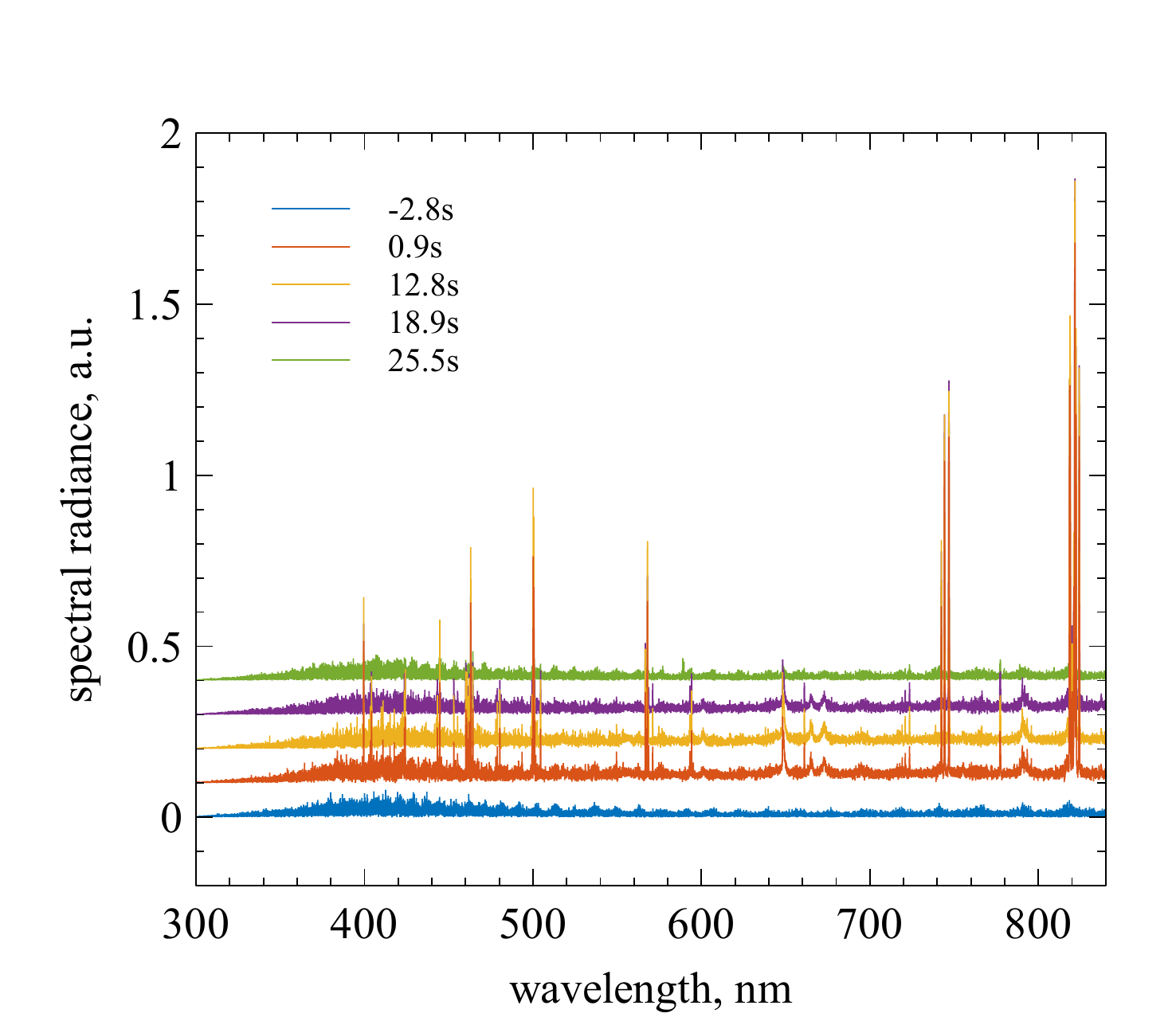}\label{fig:Al7_fl_65_Sz1_ECH_spectra}}
	\subfloat[Scenario 1 spectral radiance of the individual lines over time]{\includegraphics[width = 0.508\linewidth]{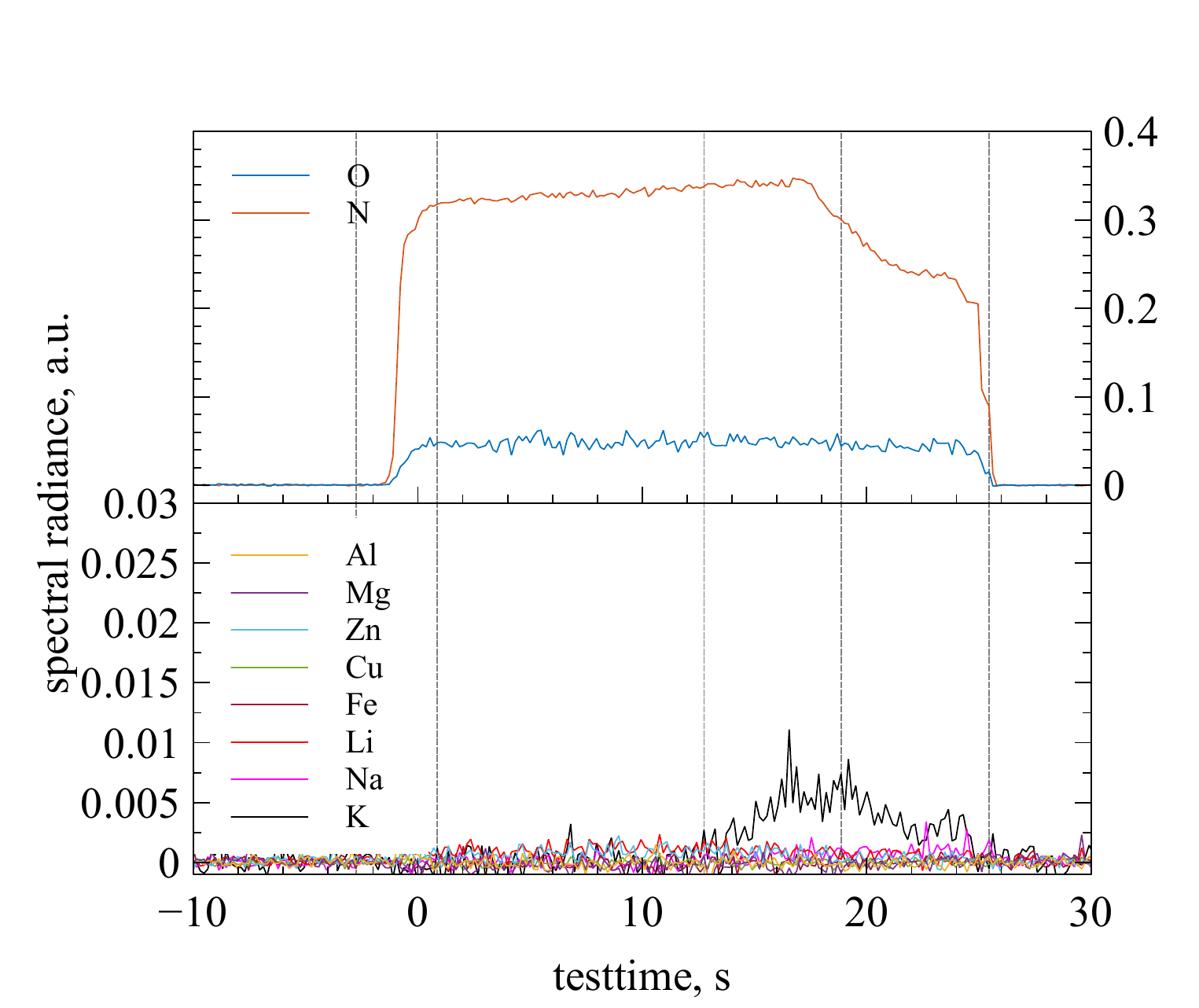}\label{fig:Al7_fl_65_Sz1_ECH_time}} \\
	\subfloat[Scenario 2 spectra at five test times of interest, marked in Fig.~\ref{fig:Al7_fl_65_Sz2_ECH_time}]{\includegraphics[width = 0.492\linewidth]{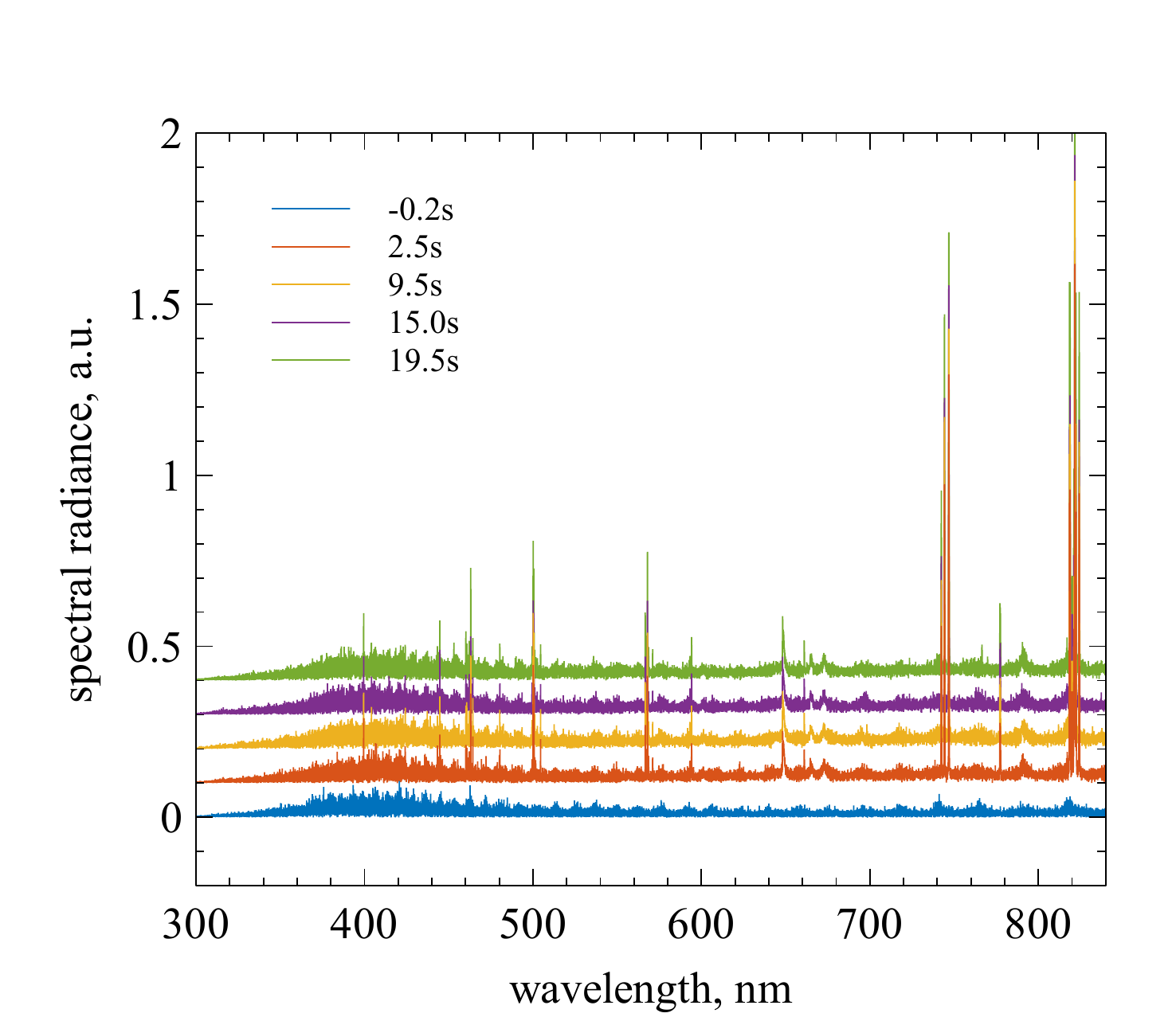}\label{fig:Al7_fl_65_Sz2_ECH_spectra}}
	\subfloat[Scenario 2 spectral radiance of the individual lines over time]{\includegraphics[width = 0.508\linewidth]{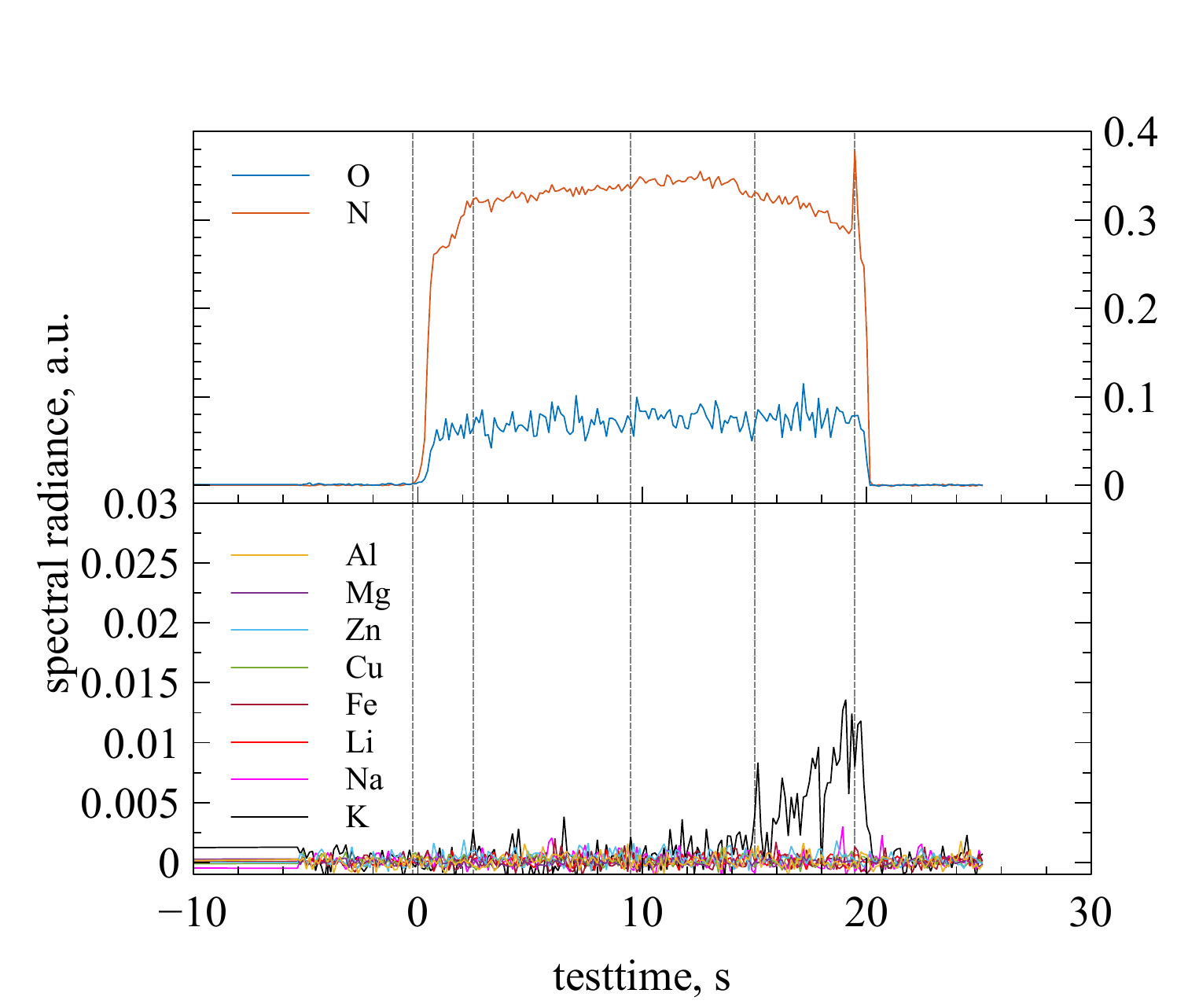}\label{fig:Al7_fl_65_Sz2_ECH_time}} 
	\caption{Echelle data of Al-7075 at 65km\label{fig:Al7_fl_65_ECH}}
\end{figure}

Data collected during the \SI{65}{\kilo\meter} experiments is shown in Fig.~\ref{fig:Al7_fl_65_ECH}.

During both tests, the most prominent lines and bands correspond to the air plasma species, most prominently \ce{N} and \ce{N+}. This is the same observation made during the Al-6060. At about \SI{14}{\second} a weak \ce{K} signature appears.
During the scenario 1 test, the \ce{K} line intensity peaks at \SI{19}{\second} and decreases toward the end of the experiment at \SI{25}{\second}.
In the scenario 2 test the experiment ends after \SI{20}{\second}, at this point the \ce{K} line intensity has peaked. The evolution of the lines is almost identical up to the point of failure.
No further elements were discernable in the spectral analysis.

\subsubsection{75km}
\begin{figure}[!ht]
	\centering
	\subfloat[Front view]{\includegraphics[trim= 45cm 0cm 45cm 0cm,clip,height = 6 cm]{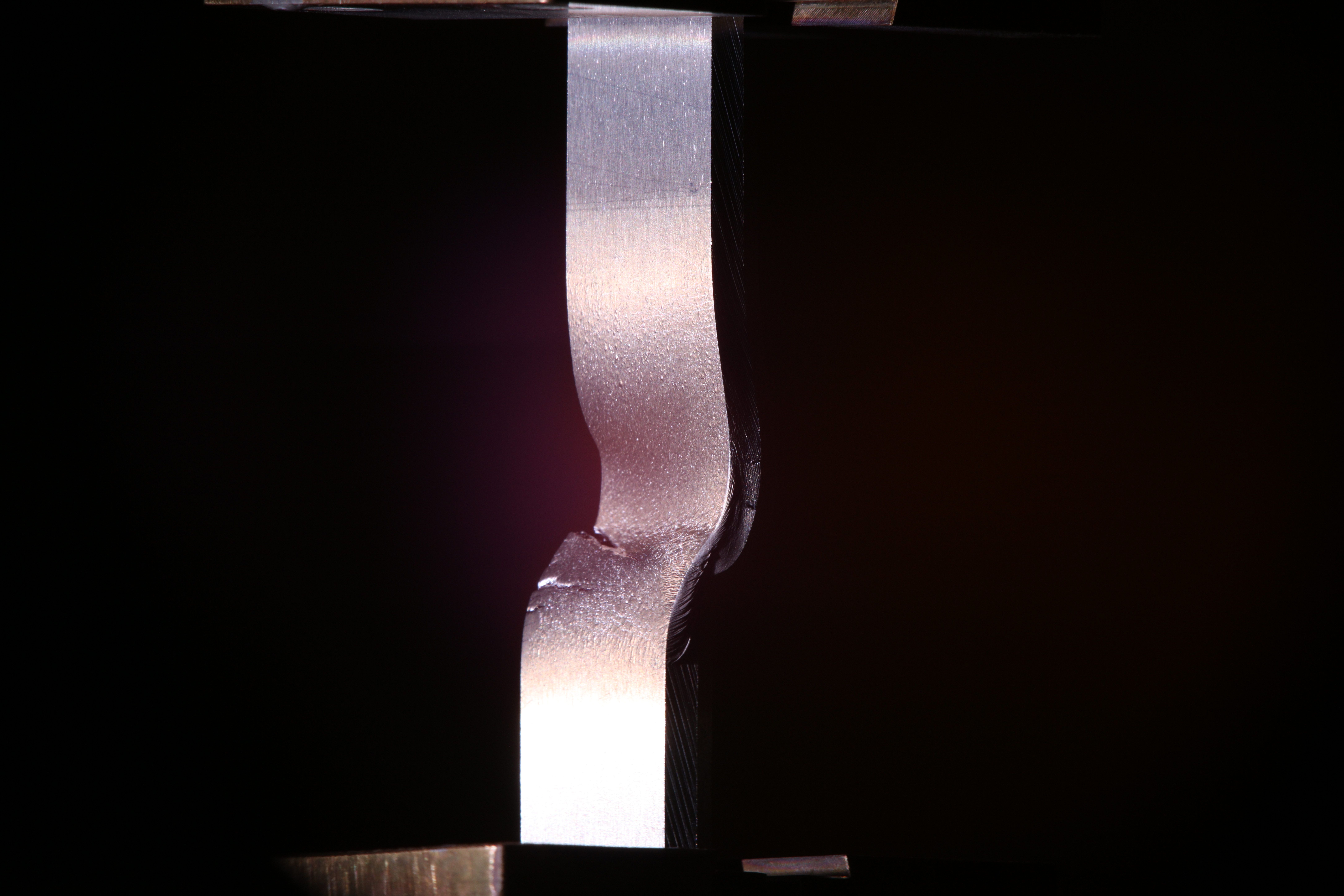}\label{fig:AL7_front}}\quad
	\subfloat[Side View]{\includegraphics[trim= 0cm 0cm 5cm 0cm,clip,height = 6cm]{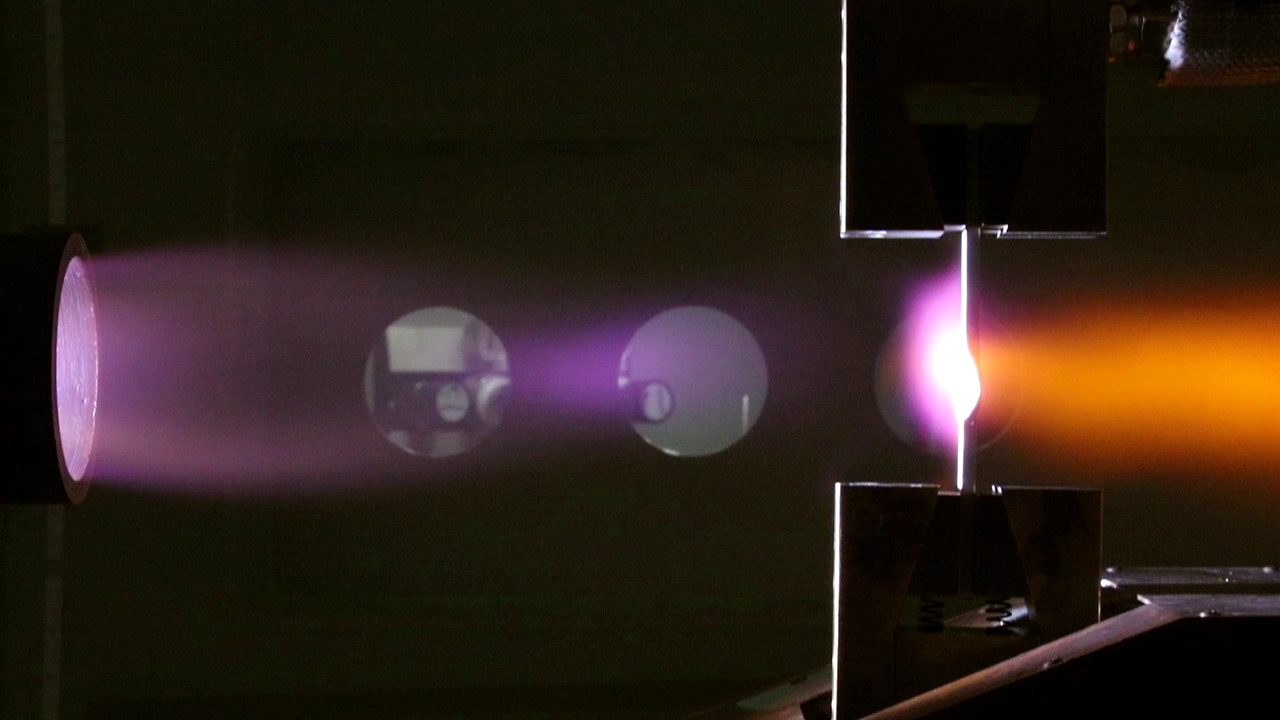}\label{fig:AL7_side}} 
	\caption{Still frame of aluminum 7075 test at 75\,km scenario 1 \label{fig:AL7_vid}}
\end{figure}

Figure~\ref{fig:AL7_vid} shows a still frame of the aluminum 7075 test at \SI{75}{\kilo\meter} with force scenario 1 shortly before the material failure. The front view (Fig.~\ref{fig:AL7_front}) shows the sample failing, it is unclear if the material is molten since the probe is deformed significantly while the surface shows cracks. An image at the same time is shown in the side view (Fig.~\ref{fig:AL7_side}). A strong discoloration is visible in both the stagnation region and the wake.

\begin{figure}[!ht]
	\centering
	\subfloat[Scenario 1 spectra at five test times of interest, marked in Fig.~\ref{fig:Al7_fl_75_Sz1_ECH_time}]{\includegraphics[width = 0.492\linewidth]{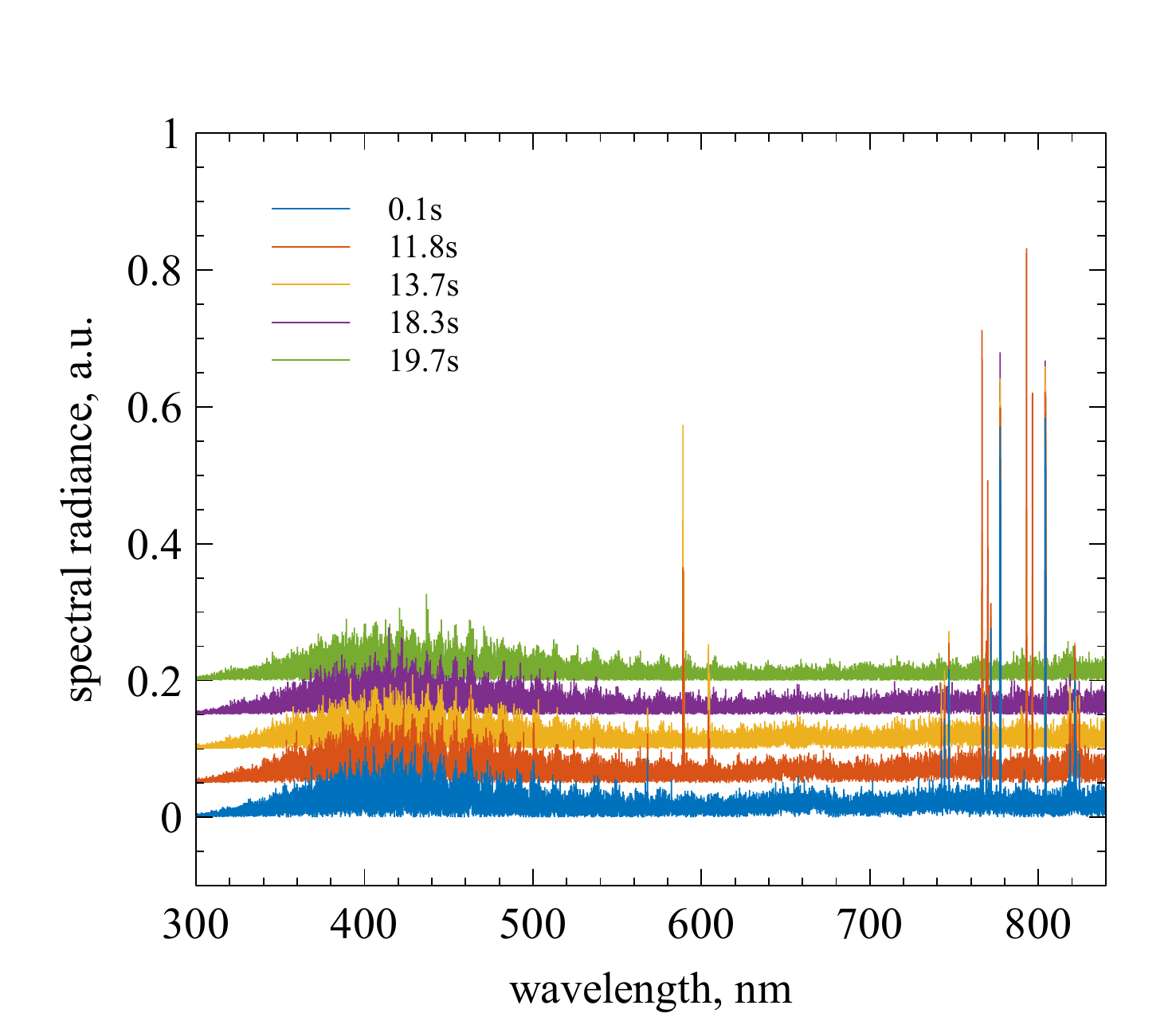}\label{fig:Al7_fl_75_Sz1_ECH_spectra}}
	\subfloat[Scenario 1 spectral radiance of the individual lines over time]{\includegraphics[width = 0.508\linewidth]{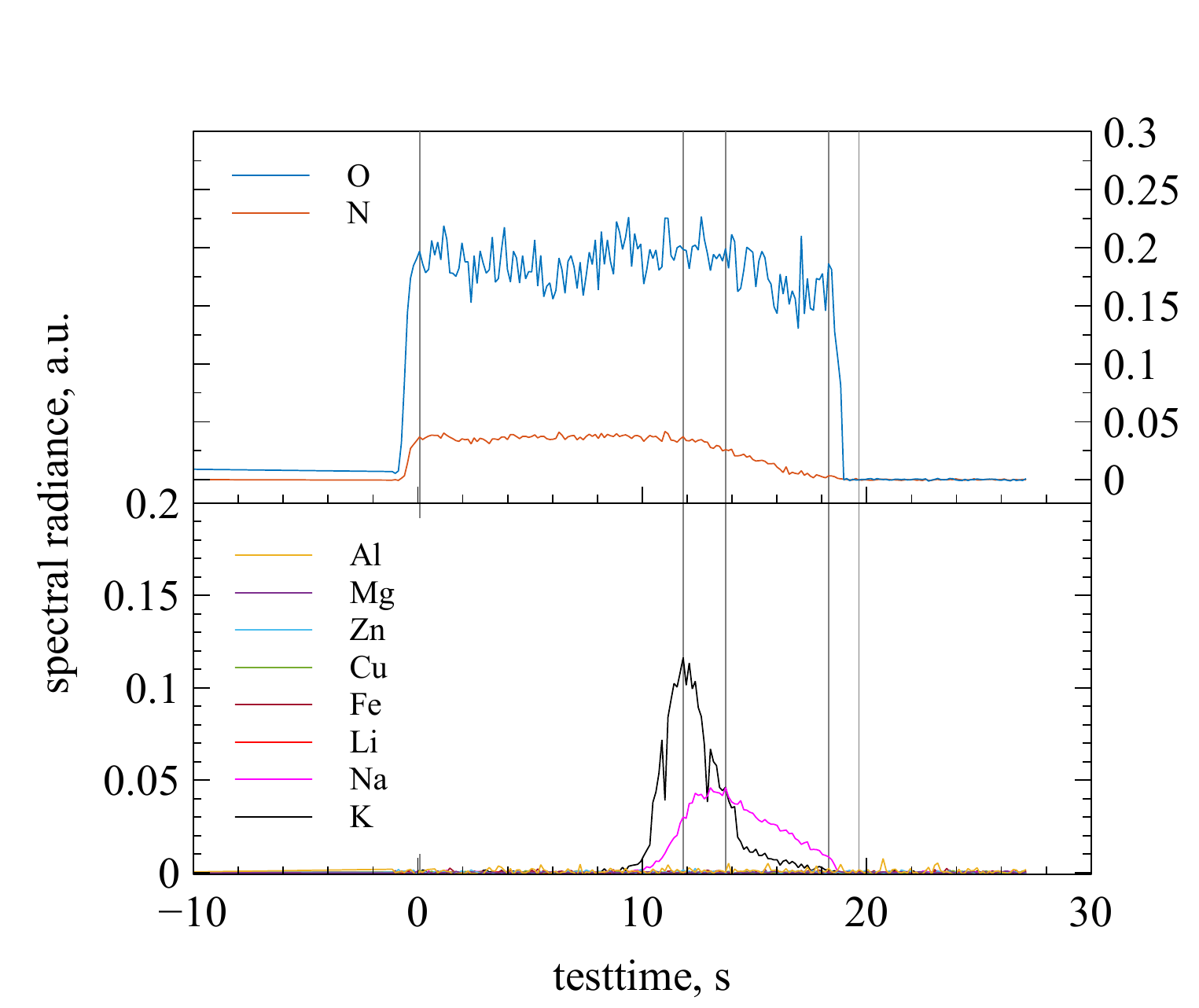}\label{fig:Al7_fl_75_Sz1_ECH_time}} \\
	\subfloat[Scenario 2 spectra at five test times of interest, marked in Fig.~\ref{fig:Al7_fl_75_Sz2_ECH_time}]{\includegraphics[width = 0.492\linewidth]{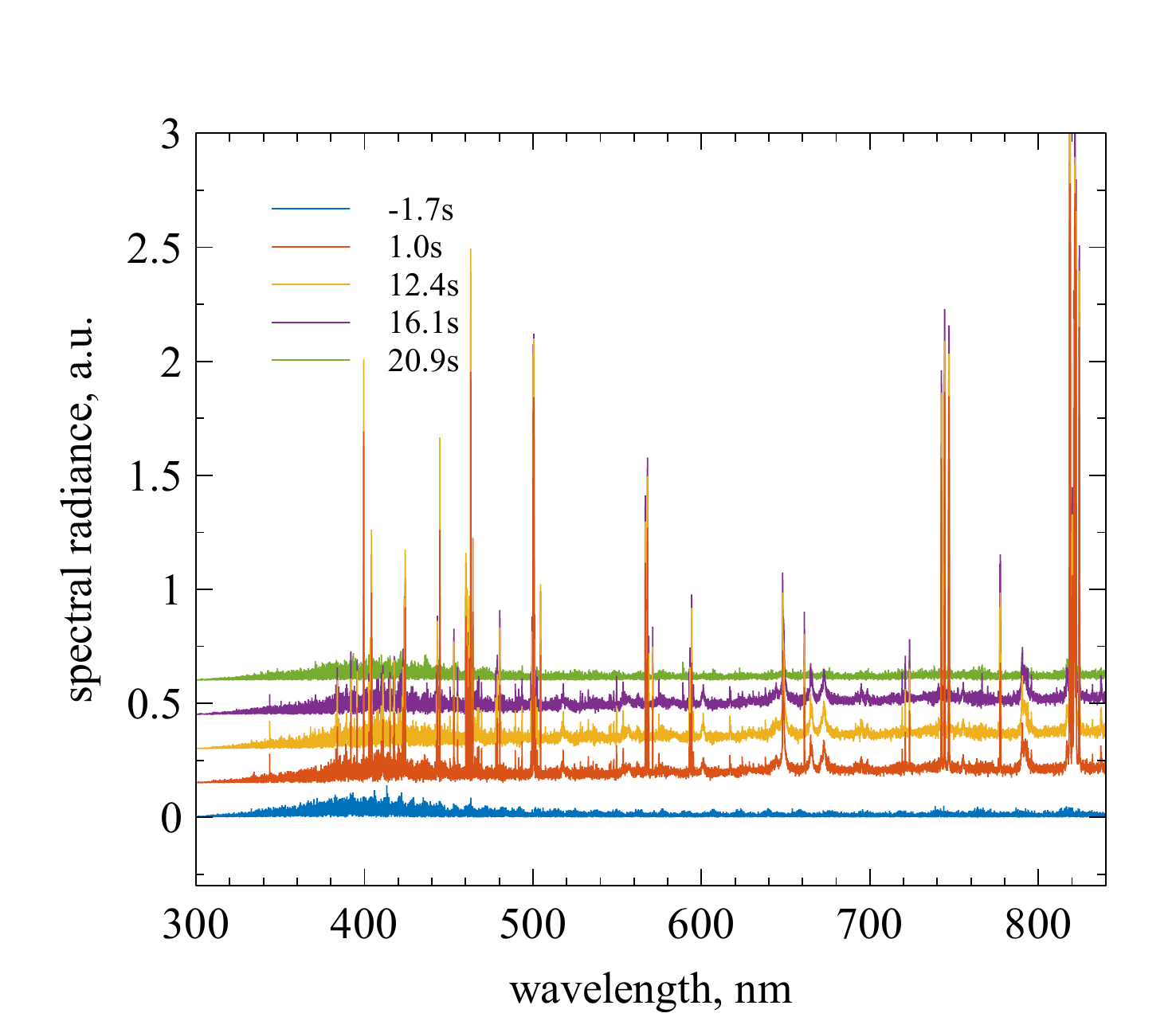}\label{fig:Al7_fl_75_Sz2_ECH_spectra}}
	\subfloat[Scenario 2 spectral radiance of the individual lines over time]{\includegraphics[width = 0.508\linewidth]{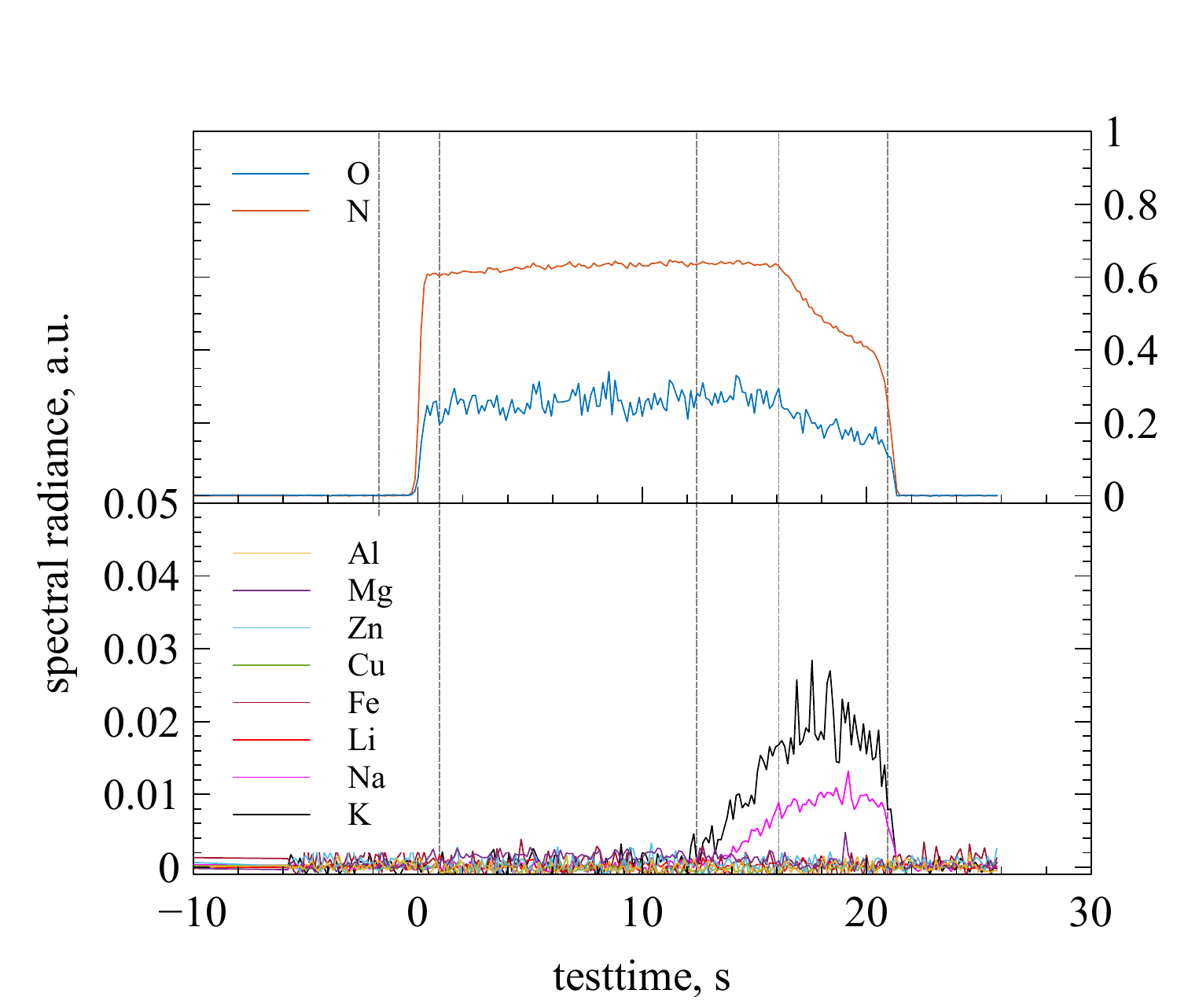}\label{fig:Al7_fl_75_Sz2_ECH_time}} 
	\caption{Echelle data of Al-7075 at 75km\label{fig:Al7_fl_75_ECH}}
\end{figure}

Spectral data of the tests at the \SI{75}{\kilo\meter} condition are displayed in Fig.~\ref{fig:Al7_fl_75_ECH}. 

The scenario 2 experiment was conducted in the latter half of the campaign, therefore the data is less noisy than at the scenario 1 condition. One obvious oddity is the fact that the sample with an applied force during the experiment (scenario 2) survived longer than without. It is assumed that this is an error in the applied condition. This explains some of the obvious discrepancies in the temporal evolution of the species.
The air plasma lines are less pronounced than in the scenario 2 test, while the alkali metals are stronger.

In the same manner as the experiments with Al-6060 aluminum none of the alloy constituents are visible spectrally. Solely the plasma lines as well as alkali metals are visible. In the tests, \ce{K} appears first at around \SI{9}{\second} and \SI{12}{\second} respectively. 
In the scenario 1 test, \ce{K} peaks at \SI{12}{\second} and weakens towards until it is no longer visible at \SI{18}{\second} , shortly before the end of the experiment at \SI{19}{\second}. In the scenario 2 test, \ce{K} peaks at \SI{18}{\second} and then weakens slightly until the end of the experiment at \SI{21}{\second}.

\ce{Na} appears at \SI{10}{\second} and \SI{14}{\second} respectively. In the scenario 1 test the evolution is very similar to that of \ce{K}, the \ce{Na} line peaks at \SI{14}{\second} and diminishes up until the test end. The same is true for the scenario 2 test, \ce{Na} peaks at \SI{19}{\second} and weakens slightly until the end of experiment.

\subsubsection{90km}
\begin{figure}[!ht]
	\centering
	\subfloat[Spectra at five test times of interest, marked in Fig.~\ref{fig:Al7_fl_90_Sz2_ECH_time}]{\includegraphics[width = 0.492\linewidth]{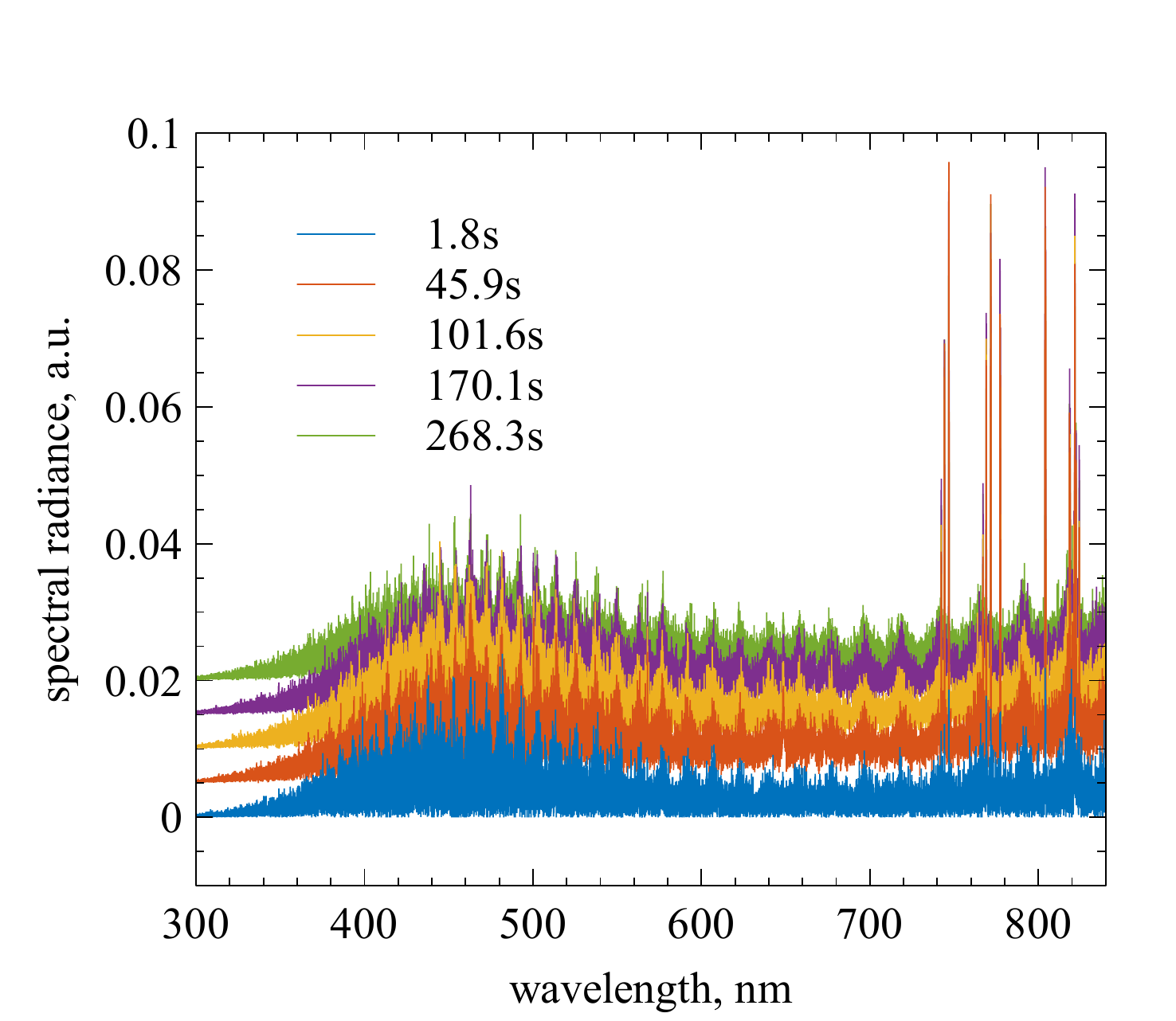}\label{fig:Al7_fl_90_Sz2_ECH_spectra}}
	\subfloat[Spectral radiance of the individual lines over time]{\includegraphics[width = 0.508\linewidth]{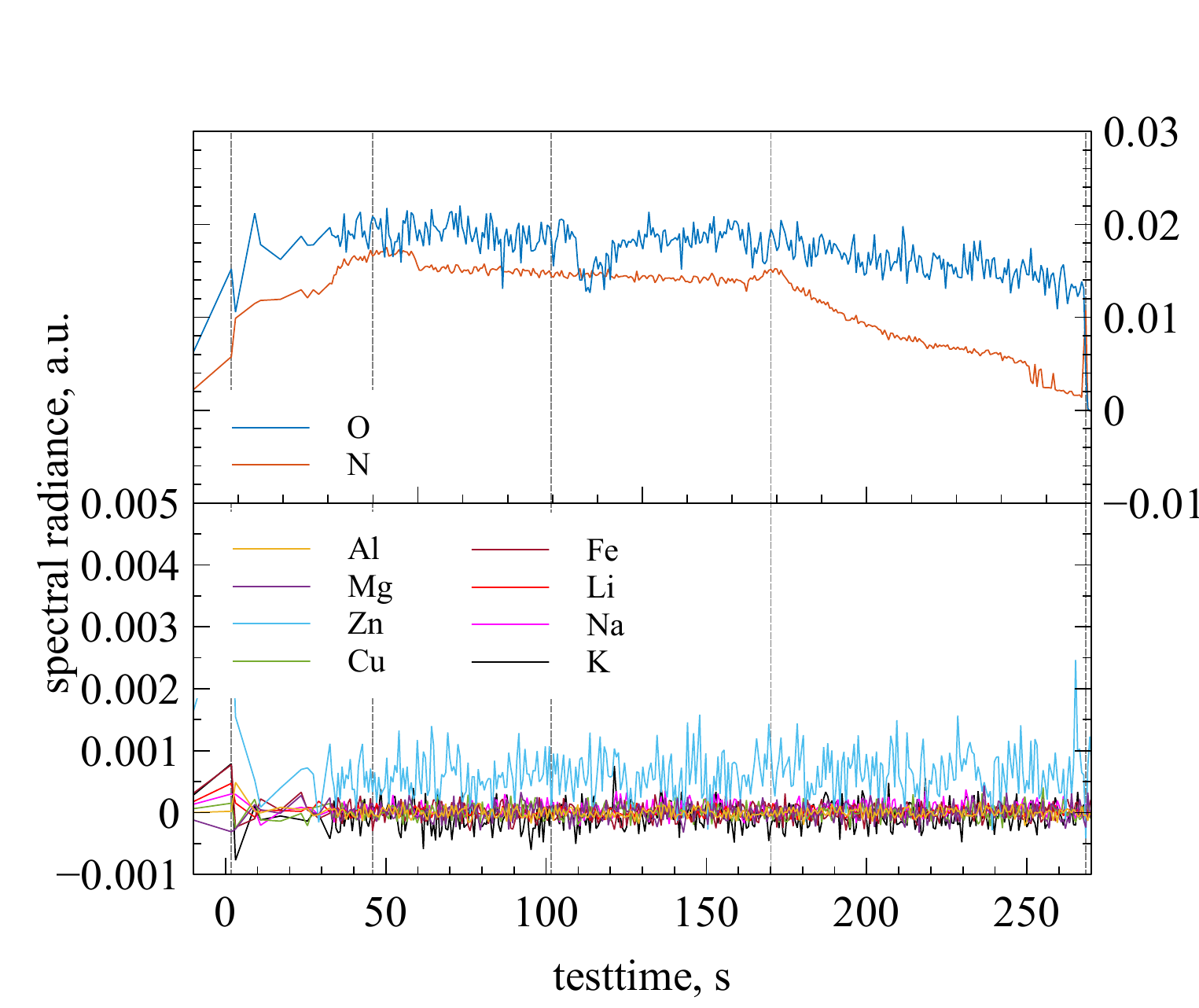}\label{fig:Al7_fl_90_Sz2_ECH_time}} 
	\caption{Echelle data of Al-7075 at 90km scenario 1/2 \label{fig:Al7_fl_90_Sz2_ECH}}
\end{figure}

Data collected during the \SI{90}{\kilo\meter} condition experiment is shown in Fig.~\ref{fig:Al7_fl_90_Sz2_ECH}. The experiment combined both the scenario 1 and 2 condition due to the sample not failing during scenario 2. The experiment was carried out in the first half of the campaign.
The condition represents a very low pressure and low heat flux, therefore the lines are very weak. No lines aside from the plasma lines were distinguishable during testing.

\section{Analysis and Comparison to Flight}
Published data on the re-entry from low earth orbit is limited, the only data known to the authors is of the observations campaigns of ATV-1~\cite{Mazoue_2010_01,Marynowski_2010_01} and Cygnus OA-6~\cite{Loehle_2017_03,Loehle_2020_10}. During the re-entry observation of the CYGNUS OA-6 spacecraft, sodium and lithium were detected. During the ATV re-entry observation, the presence of lithium was detected from the structural break-up explosion onward and peaked later into the entry. This was assumed to be a result of the lithium-ion (\ce{LiMnO2}) batteries~\cite{Lagattu_2011_01} being damaged. However, the fact that manganese lines appeared later speaks to the contrary~\cite{Lips_2011_01}. 

At the \SI{90}{\kilo\meter} test, only spectral lines attributed to the air plasma are visible. No significant changes occur and no other lines have been identified. This behaviour fits to the re-entry simulations and observations in that no major structural material failure occurs at these altitudes. 

At the \SI{75}{\kilo\meter} condition the material failure was always preceded by the emission of alkali metals, regardless of the applied strength and the resulting variation in failure mode. A very weak potassium signature is visible in the last \SIrange{5}{10}{\second} of all tests at \SI{75}{\kilo\meter}. The emission of sodium at different strengths is also visible in all experiments, the emission of lithium is specific to the Al-6060  alloy and dominant in the case with no force and a melting sample. 

In comparison to the observed and simulated re-entries, \SI{75}{\kilo\meter} is equivalent to the break-up altitude of spacecraft entering from low-earth orbit. A strong onset of alkali metal emission may be used to better determine the time and thus the altitude of the main spacecraft break-up. 

The \SI{65}{\kilo\meter} condition shows similar features as the \SI{75}{\kilo\meter} condition, all spectra show the emission of potassium shortly before failure. However the emission of sodium is particular to Al-6060 and the emission of lithium is visible upon the material melting. As the condition that is representative of the altitude that features internal component exposure such as reaction wheels, batteries and electronic components, as well as being the trajectory point of peak heating many species are visible in the spectra of re-entry observations. It was therefore expected that more alloying elements would be visible in the experiments. The authors assume that the presence of alloying or base elements is a sign of material demise rather than structural break-up.

The new insights collected during this analysis speaks to the fact that the presence of alkali metal emission is due to the material failure of structural components made of aluminum alloys. %The fact that no alloying elements were visible in any of the test, they are observed through large portions of re-entry observations could be smaller components broken off that are exposed to much higher heat fluxes and melt more rapidly. 
A major difference between the test and re-entry is the fact that no alloying elements were visible in any of the test, while they are observed through large portions of re-entry observations. One possibility is that alloying elements only appear later in the demise process. Smaller components that have broken off are exposed to a higher heat flux, this expedites melting and dispersion to smaller particles.
Since all experiments were ended upon the sample failure no information was obtained on the spectral response of demising aluminum alloys.

These observations show that even though the alkali metals, potassium, sodium, and lithium are not alloying elements of the aluminum alloys Al-6060 and Al-7075 they appear in direct correlation with the material failure. While potassium and sodium appear throughout the materials and conditions lithium is only visible in experiments featuring Al-6060 and most prominently upon the bulk material melting. 

The authors therefore recommend to review the observed alkali metal emission in past observations and focus for these features in future re-entry observation missions. The emission of lithium from battery failure may be easily separated from the structural failure of aluminum structures and give more insight into the degradation of spacecraft aluminum structures.

\section{Conclusion}
Aluminum alloys were tested at conditions representative for typical flight paths. The three conditions correspond to trajectory points at early entry, break-up, and demise. 
The materials were tested under combined thermochemical and aeromechanical loads in a plasma wind tunnel facility with applied mechanical forces.
The first interesting feature is that the analysis of emission spectra recorded with an Echelle spectrometer shows no emission of alloying elements. However strong signals corresponding to the alkali metals lithium, sodium, and potassium were detected. This is the second main result: These emissions occur shortly before the samples failed. It stands to reason that the emission of these elements can be used as an indicator of the failure of these materials during re-entry observation missions. 
The third finding is that a comparison of these ground tests to spectra recorded during re-entry of Cygnus OA-6 \emph{S.S. Rick Husband} and ATV-1 \emph{Jules Verne} show an earlier than expected emission of sodium and lithium in particular. Lithium was specifically attributed to the battery systems while sodium was thought to have originated from human waste. The new findings during this analysis might lead to a different interpretation. Future re-entry observation missions should therefore focus on alkali metals as indicators of material failure.
As the alkali metals are strong emitters, trace amounts may already be detected early on. While the bulk of the spacecraft is intact, the observation of alkali metals is most likely only from structural materials. After the main fragmentation event, additional care must be taken to separate possible sources. In the case of lithium-ion (\ce{LiMnO2}) batteries, the detection of manganese could be a direct indicator of battery failure.
The fact that no alloying elements were detected in ground testing, but that strong signals were recorded during re-entry observations points to these elements being released during the demise of aluminum components. %Future investigations into the spectral characteristics of demising structures are therefore a logical next step.
This paper concludes that fragmentation of aluminum can be detected by alkali metal line emission. 

%\section*{Appendix}

\section*{Acknowledgments}
The authors gratefully acknowledge the financial support by the German Aerospace Center (DLR) through the research grant No. 50LZ1704. The dedicated support of the institute’s workshop is indispensable and gratefully acknowledged. The authors would like to thank the colleagues from the High Enthalpy Flow Diagnostics Group for their continuous support.

\bibliography{literatur_Merkliste}

\end{document}